\documentclass[12pt]{article}
\usepackage{graphicx}
\usepackage{epstopdf}
\usepackage{amsmath}
\usepackage{bm}
\usepackage{multirow}
\usepackage{graphicx}
\usepackage{caption}
\usepackage{subcaption}
\usepackage{rotating}
\usepackage{array}
\usepackage{amsfonts}
\usepackage{afterpage}
\def\lsim{\mathrel{\raise.3ex\hbox{$<$\kern-.75em\lower 1ex\hbox{$\sim$}}}}
\def\gsim{\mathrel{\raise.3ex\hbox{$>$\kern-.75em\lower 1ex\hbox{$\sim$}}}}

\def\be{\begin{equation}}
\def\ee{\end{equation}}
\def\bea{\begin{eqnarray*}}
\def\eea{\end{eqnarray*}}

\begin{document}
\title{Neutrino seesaw mechanism with texture zeros}
\author{Jiajun Liao$^{1,3}$, D. Marfatia$^{1,3}$, and K. Whisnant$^{2,3}$\\
\\
\small\it $^1$Department of Physics and Astronomy, University of Hawaii at Manoa, Honolulu, HI 96822, USA\\
\small\it $^2$Department of Physics and Astronomy, Iowa State University, Ames, IA 50011, USA\\
\small\it $^3$Kavli Institute for Theoretical Physics, University of California, Santa Barbara, CA 93106, USA}
\date{}
\maketitle

\begin{abstract}
In the context of the Type I seesaw mechanism, we carry out a
systematic study of the constraints that result from zeros in both the
Dirac and right-handed Majorana neutrino mass matrices. We find that
most constraints can be expressed in the standard form with one or two
element/cofactor zeros alone, while there are 9 classes of
{\it nonstandard} constraints. We show that all the constraints are stable under one-loop renormalization group
running from the lightest right-handed neutrino mass scale to the
electroweak scale. We study the
predictions of the nonstandard constraints for the lightest neutrino mass, Dirac CP phase and neutrinoless
double beta decay. 
\end{abstract}

\newpage

\section{Introduction}
Many Standard Model (SM) extensions predict the existence of right-handed (RH)
Majorana neutrinos, which are often used to explain neutrino masses via the
Type I seesaw mechanism~\cite{seesaw}. The effective
mass matrix of the light Majorana neutrinos is given by
\begin{align}
M=-M_D {M_R}^{-1}M_D^T\,,
\label{eq:Mseesaw}
\end{align}
where $M_D$ and $M_R$ are the mass matrices for the Dirac and RH Majorana neutrinos, respectively. 

In the basis where the charged lepton mass matrix is diagonal, the
light neutrino mass matrix is described by 9 real parameters and can
be written as~\cite{Agashe:2014kda}
\begin{equation}
M=V^*\text{diag}(m_1, m_2, m_3)V^\dagger\,,
\label{eq:Mnu}
\end{equation}
where the $m_i$ are real and nonnegative, $V=U\text{diag}(1, e^{i\phi_2/2}, e^{i\phi_3/2})$, and 
\begin{align}
U=\begin{bmatrix}
   c_{13}c_{12} & c_{13}s_{12} & s_{13}e^{-i\delta} \\
   -s_{12}c_{23}-c_{12}s_{23}s_{13}e^{i\delta} & c_{12}c_{23}-s_{12}s_{23}s_{13}e^{i\delta} & s_{23}c_{13} \\
   s_{12}s_{23}-c_{12}c_{23}s_{13}e^{i\delta} & -c_{12}s_{23}-s_{12}c_{23}s_{13}e^{i\delta} & c_{23}c_{13}
   \end{bmatrix}\,.
\label{eq:U}
\end{align}

After decades of neutrino oscillation experiments,
five parameters (three mixing angles and two mass-squared differences)
in the neutrino sector have been determined to rather good
precision. However, the hierarchy of the three light neutrino masses
is still unknown, and there are two possibilities: \textit{(i)}
$m_1<m_2<m_3$ for the normal hierarchy (NH) and \textit{(ii)}
$m_3<m_1<m_2$ for the inverted hierarchy (IH). The remaining four
unknown parameters may be taken as the lightest mass ($m_1$ for the
NH, or $m_3$ for the IH), the Dirac CP phase $\delta$ and two Majorana
phases ($\phi_1$ and $\phi_2$). The Dirac phase and the mass hierarchy
will be measured in future neutrino oscillation experiments, and the
lightest mass can in principle be determined from beta decay experiments and
cosmological observations.

Of the many neutrino mixing models proposed in the literature, texture
zero models~\cite{Frampton:2002yf} are attractive because of
the simple relations on the matrix elements that are found in these models. Texture zero
constraints can be imposed on the light neutrino mass matrix
directly. However, in the context of the Type I seesaw mechanism,
since the light mass matrix is a product of the Dirac mass matrix and
the RH neutrino mass matrix, it is natural to consider texture
zeros in the Dirac and RH neutrino mass
matrices~\cite{Lavoura:2015wwa}. Zeros in the Dirac and RH neutrino
mass matrix can be realized from discrete $\mathbb{Z}_N$ symmetries
with suitable scalar singlets~\cite{Grimus:2004hf}.

In this paper, we carry out a systematic study of the constraints that
result from various zero textures in both the Dirac and RH Majorana
mass matrices in the context of the Type I seesaw mechanism. We find
that most cases lead to the standard constraints which can be expressed in
the form of just one or two element/cofactor zeros, which have been
studied extensively in the literature~\cite{Lashin:2011dn, Liao:2013kix, Liao:2013rca, Liao:2013saa, TT, CC, TC}. However, there are also 9
classes of {\it nonstandard} constraints, i.e., constraints which are not
described by simple texture or cofactor zeros in the light neutrino
mass matrix. We first show that both the
standard and nonstandard constraints are stable under one-loop
renormalization group equation (RGE) running from the lightest RH
neutrino mass scale $M_1$ to the electroweak scale $M_Z$. Then we
study the phenomenological implications of the nonstandard
constraints using recent data measured by neutrino oscillation
experiments. We find that some cases
are excluded, and for the rest we obtain preferred values for the
lightest mass and Dirac CP phase, which will be probed in the next
generation of neutrino experiments. We also study neutrinoless
double beta decay for the nonstandard constraints.

This paper is organized as follows. In Section~2, we explore the constraints of zero textures in the seesaw mechanism. In Section~3, we discuss RGE effects on the constraints. In Section~4, we study the phenomenological implications of the nonstandard constraints and summarize our results in Section~5.

\section{Classes of constraints}
We consider three generations of RH Majorana neutrinos and we
assume $M_R$ is not singular. Since the Majorana mass matrix is
symmetric, there are 6 independent matrix elements in $M_R$, and 9
independent elements in $M_D$. For a case with $N_R$ zeros in $M_R$
and $N_D$ zeros in $M_D$, there are should be at least $N-6$ complex
constraints on the elements of $M$, where $N=N_R+N_D$ is the total
number of zeros element. This may be understood as follows.

We start with $15=6+9$ nonzero complex parameters in $M_R$ and $M_D$,
respectively. There are a total of $N$ zeros in $M_R$ and $M_D$, so that
leaves $15 - N$ independent complex parameters. We can always insert a diagonal matrix between $M_R$ and $M_D$ (and the same matrix
between $M_R$ and $M_D^T$), which effectively absorbs 3 complex
parameters (or, putting it another way, there are 3 relative
normalizations that are redundant in the prediction of the light
neutrino mass matrix from the seesaw mechanism). This leaves us with
12 - N independent complex parameters. In $M$ there are 6
independent elements, so the net number of constraints on the elements
of $M$ is $6 - (12 - N) = N - 6$. In terms of real parameters, we have $24 - 2 N$, from which we
can subtract 3 unphysical phases, leaving us with $21 - 2 N$ physical
real parameters in $M$. Since there are 9 real parameters in the
neutrino sector and 5 of them are well determined, in general, if
$N\geq 9$, the model will be over constrained and if $N\leq 6$, the
model will be unconstrained.

This counting rule only gives the minimum number of constraints. There
are accidental constraints that result from some special combinations
of texture zeros in $M_D$ and $M_R$, e.g., if $M_D$ has a column or a
submatrix composed of all zeros, then the constraint $\det M=0$ always
exists regardless of the counting rule. So in order to obtain all the
constraints for a given zero structure of $M_D$ and $M_R$, we have to
make sure all of the independent constraints are obtained. Our analysis
proceeds as follows.

We scan all possible texture zero structures for $M_D$. Since $M$ is
invariant under a permutation of the columns and rows of $M_R$ with a
simultaneous permutation of the corresponding columns of $M_D$, we
only consider one representative structure for each set of $M_R$ that
are equivalent under permutation. For each combination of $M_D$ and
$M_R$, we check all the elements and cofactors, and the determinant of $M$
using Eq.~(\ref{eq:Mseesaw}) and look for independent constraints. We
write the constraints in terms of the elements, cofactors or
determinant of $M$ in the simplest form. The experimental data does
not allow more than two zeros in the elements or cofactors of the
light mass matrix, hence we only consider cases that give less
than three such zeros. The cases that give a block diagonal mass matrix are also
ignored because they lead to a zero mixing angle, which is excluded by
current experimental data. For the other cases, we analyze the
elements of $M$, and make sure we obtain all of the independent
constraints. We find that most constraints can be expressed in the
standard form that can be written in terms of just one or two
element/cofactor zeros, but there are also nine nonstandard constraints,
which are listed in Table~\ref{tb:constraints}. The structures that
lead to the nonstandard constraints are provided in
Appendix~\ref{ap:structure}. Classes 2, 3, 4 and 5 are discussed in
Ref.~\cite{Lavoura:2015wwa}, with constraint equations in a different (albeit equivalent) form. Classes 1 and 9 are subsets of
well-known models with the additional constraint $\det M = 0$. Classes
6, 7 and 8 are new and not previously discussed in the literature.

\begin{table}[t]
\caption{Classes of the nonstandard constraints that can not be
expressed in terms of one or two element ($M_{\alpha\beta}$) or
cofactor ($C_{\alpha\beta}$) zeros alone. Here $\alpha \neq \beta
\neq \gamma$, and no sum is implied in the constraint equations.}
\begin{center}
\begin{tabular}{|c|c|c|c|}\hline
Class & $N_R+N_D$ & Nonstandard constraints  & No. of constraints \\ \hline
1 & $4+5$ & $M_{\alpha\alpha}=M_{\beta\beta}=0$ and $\det M=0$ & 3 \\\hline
2 & $4+4$, $3+5$, $3+4$, $2+5$  & $M_{\alpha\alpha}=0$ and $\det M=0$ & 3 \\\hline
3 & $3+5$ & $M_{\alpha\beta}=0$ and $\det M=0$ & 3 \\\hline
4 & $4+4$ & $M_{\alpha\alpha}=0$ and $M_{\beta\beta}M_{\alpha\gamma}=2M_{\alpha\beta}M_{\beta\gamma}$ & 6\\\hline
5 & $3+5$ & $C_{\alpha\alpha}=0$ and $C_{\beta\beta}C_{\alpha\gamma}=2C_{\alpha\beta}C_{\beta\gamma}$ & 6 \\\hline
6 & $4+3$, $3+4$, $2+5$  & $M_{\alpha\alpha}C_{\alpha\alpha}=\det M$ & 3 \\\hline
7 & $4+3$ & $M_{\alpha\alpha}^2C_{\alpha\alpha}=4M_{\alpha\beta}M_{\alpha\gamma}C_{\beta\gamma}$ & 3 \\\hline
8 & $3+4$ & $M_{\beta\beta}\det M =-M_{\alpha\beta}^2C_{\alpha\alpha}$ & 6 \\\hline
9 & $4+3$, $3+4$, $2+5$ & $\det M=0$ & 1 \\\hline
\end{tabular}
\end{center}
\label{tb:constraints}
\end{table}

\section{Renormalization group running}
The neutrino mixing parameters are measured at low energies, while the effective mass matrix of
Eq.~(\ref{eq:Mseesaw}) can arise at a much high energy scale.  The
evolution of the light neutrino masses from the lightest RH neutrino
mass scale $M_1$ to the electroweak scale $M_Z$ is described by the
one-loop RGE~\cite{Antusch:2005gp},
\begin{equation}
16\pi^2\frac{dM}{dt}=\kappa M + \eta[(Y_lY_l^\dagger)M+M(Y_lY_l^\dagger)^T]\,,
\label{eq:RGE}
\end{equation}
where $t = \ln(\Lambda/M_1)$, $\Lambda$ is the renormalization scale and
$Y_l=\text{diag}(y_e, y_\mu, y_\tau)$ with $y_\alpha$ being the eigenvalues of the charged lepton Yukawa
coupling matrix. In the SM, 
\begin{align}
\eta=-\frac{3}{2}\,, \quad \kappa\approx-3g_2^2+6y_t^2+\lambda\,,
\end{align}
and in the Minimal Supersymmetric Standard Model (MSSM), 
\begin{align}
\eta=1\,, \quad \kappa\approx-\frac{6}{5}g_1^2-6g_2^2+6y_t^2\,,
\end{align}
where $g_1,g_2$ are
the gauge couplings, $y_t$ is the top quark Yukawa coupling, and
$\lambda$ is the Higgs self-coupling. 

The solution to Eq.~(\ref{eq:RGE}) can be written in the form~\cite{solution},
\begin{equation}
M(M_Z)=I_0\begin{bmatrix}
   I_e & 0 & 0 \\
   0 & I_\mu & 0  \\
   0 & 0 & I_\tau
   \end{bmatrix}M(M_1)\begin{bmatrix}
      I_e & 0 & 0 \\
      0 & I_\mu & 0  \\
      0 & 0 & I_\tau
      \end{bmatrix},
\label{eq:RGEsolution}
\end{equation}
where 
\begin{equation}
I_0=\exp\left[-\frac{1}{16\pi^2}\int_0^{\ln(M_1/M_Z)}\kappa(t)dt\right],
\end{equation}
and 
\begin{equation}
I_l=\exp\left[-\frac{\eta}{16\pi^2}\int_0^{\ln(M_1/M_Z)}y_l^2(t)dt\right],
\end{equation}
for $l=e,\mu,\tau$. 

From Eq.~(\ref{eq:RGEsolution}), we see that RGE running can be described by multiplying the elements of the mass matrix by three factors, i.e.,
\begin{align}
M_{\alpha\beta}= M'_{\alpha\beta} I_0 I_\alpha I_\beta\,,
\end{align}
where $\alpha,\beta=e,\mu,\tau$, $M$ and $M'$ are the effective mass matrix at the electroweak scale and the lightest RH neutrino mass
scale, respectively. Taking the inverse of Eq.~(\ref{eq:RGEsolution}), we get
\begin{align}
(M^{-1})_{\alpha\beta}= (M'^{-1})_{\alpha\beta} \frac{1}{I_0I_\alpha I_\beta}\,.
\end{align}
Since $C_{\alpha\beta}=\det M (M^{-1})_{\alpha\beta}$, it is easy to verify that both the standard and nonstandard constraints are stable under one-loop renormalization group running from $M_1$ to $M_Z$.

Note that above the lightest RH neutrino mass, one must consider seesaw threshold effects, which can lead to large corrections to $\theta_{12}$ and the Dirac and Majorana phases~\cite{Antusch:2005gp}. Due to these effects, some textures that are excluded by 
data may be allowed above the seesaw threshold~\cite{Hagedorn:2004ba}. Estimating the effect of RGE running above $M_1$ on low-energy parameters is beyond the scope of this paper because additional assumptions about the Yukawa coupling matrix and the RH Majorana mass matrix need to be invoked~\cite{Ohlsson:2013xva}.

It has been shown that the two-loop RGE can change the rank of the light neutrino mass matrix~\cite{Petcov:1984nz}, which implies that the constraint $\det M=0$ is not stable at two loop. In the SM, two-loop effects can be described by adding an additional term to Eq.~(\ref{eq:RGE})~\cite{Davidson:2006tg}, so that
\begin{equation}
16\pi^2\frac{dM}{dt}=\kappa M + \eta[(Y_lY_l^\dagger)M+M(Y_lY_l^\dagger)^T]+\frac{2}{16\pi^2}(Y_lY_l^\dagger)M(Y_lY_l^\dagger)^T\,.
\end{equation}
The solution to the two-loop RGE is~\cite{Ray:2010fa}
\begin{equation}
M(M_Z)=I_0\begin{bmatrix}
   M'_{ee}I_e^2I_{ee} & M'_{e\mu}I_eI_\mu I_{e\mu} & M'_{e\tau}I_eI_\tau I_{e\tau} \\
   M'_{e\mu}I_eI_\mu I_{e\mu} & M'_{\mu\mu}I_\mu^2I_{\mu\mu} & M'_{\mu\tau}I_\mu I_\tau I_{\mu\tau}  \\
   M'_{e\tau}I_eI_\tau I_{e\tau} & M'_{\mu\tau}I_\mu I_\tau I_{\mu\tau} & M'_{\tau\tau}I_\tau^2I_{\tau\tau}
   \end{bmatrix},
\end{equation}
where 
\begin{equation}
I_{mn}=\exp\left[-\frac{2}{(16\pi^2)^2}\int_0^{\ln(M_1/M_Z)}y_m^2(t)y_n^2(t)dt\right],
\end{equation}
for $m,n=e,\mu,\tau$. We see that the nonstandard constraints are not stable under two-loop running. However, as shown in Ref.~\cite{Davidson:2006tg}, the two-loop correction to the lightest mass is of order $10^{-13}$ eV in the SM and $10^{-10}(\tan \beta/10)^4$ eV in the MSSM. Two-loop effects can also generate a tiny value for $\theta_{13}$ at the level of $10^{-12}-10^{-14}$ in the SM~\cite{Ray:2010fa}. Considering current uncertainties in the oscillation parameters, two-loop corrections do not affect our numerical results.

\section{Phenomenology of the nonstandard constraints}
The phenomenology of the standard constraints have been studied extensively in the literature; for recent analyses, see Ref.~\cite{Lashin:2011dn} for one element zero, Ref.~\cite{Liao:2013kix} for one off-diagonal element/cofactor zero, and Ref.~\cite{Liao:2013rca} for one diagonal element/cofactor zero. The results for two element/cofactor zeros can be found in Ref.~\cite{Liao:2013saa}; other recent analyses can be found in Ref.~\cite{TT} for two element zeros, Ref.~\cite{CC} for two cofactor zeros, and Ref.~\cite{TC} for one element zero and one cofactor zero. Here we focus on the phenomenology of the nonstandard constraints. 

\subsection{Classes 1, 2, 3, 9: det {\it M} = 0}

Structures with the constraint $\det M=0$ are easiest to
analyze. Since $\det M=0$ is equivalent to $m_1=0$ ($m_3=0$) for the
NH (IH), the oscillation parameters are not affected, and
these models are the same as models that have already been studied,
with the additional constraint that the lightest mass is zero. For
Class 9, $\det M = 0$ is the only constraint and therefore any value
of $\delta$ is allowed (see Sec.~5 for a discussion of the possible
constraints from neutrinoless double beta decay).
For models with constraints besides $\det M=0$, we only
need to check the parameter space allowed by the additional constraints and
see if $m_1=0$ ($m_3=0$) is allowed for the NH (IH).

Class 1 constraints have the same number of parameters as the most
economical seesaw model~\cite{Frampton:2002qc}. From Ref.~\cite{Liao:2013saa},
we know that if two diagonal elements are both equal to zero, only
$M_{\mu\mu}=M_{\tau\tau}=0$ for the IH is allowed, but in this case the
$2\sigma$ allowed region does not include $m_3=0$. Hence, Class 1 is not allowed by the experimental data at
$2\sigma$.

For Class 2, from Ref.~\cite{Liao:2013rca} we know that only
$M_{\mu\mu}=0$ or $M_{\tau\tau}=0$ are allowed for the IH at $2\sigma$
when the lightest mass is zero. Similarly, for Class 3, from
Ref.~\cite{Liao:2013kix} we know that only $M_{e\mu}=0$ or
$M_{e\tau}=0$ are allowed for the IH at $2\sigma$ when the lightest
mass is zero. In all other cases with exactly one texture zero in the
light neutrino mass matrix, a massless neutrino is
excluded at $2\sigma$. Hence only four cases
in Class 2 and 3 are allowed for the IH at $2\sigma$, and none for the
NH. The allowed cases in Class 2 are: \textit{(i)} $M_{\mu\mu}=0$, and
$\det M=0$; \textit{(ii)} $M_{\tau\tau}=0$, and $\det M=0$. The
allowed cases in Class 3 are: \textit{(i)} $M_{e\mu}=0$, and $\det
M=0$; \textit{(ii)} $M_{e\tau}=0$, and $\det M=0$. We note that the
Class 2 and 3 constraints have been noted in Ref.~\cite{Lavoura:2015wwa} as Eqs. (48) and (49). Our results for the allowed cases are
consistent with Ref.~\cite{Lavoura:2015wwa}.

Reference~\cite{Lavoura:2015wwa} also lists the conditions
$\det M=0$ and $M_{\alpha\alpha}M_{\beta\beta}-(M_{\alpha\beta})^2=0$ (as Eq.~50)
and $M_{\alpha\alpha}=M_{\alpha\beta}=0$ and
$(M^{-1})_{\alpha\alpha}=0$ (as Eq.~51) as new constraints. The
structures that lead to these conditions have six and three cofactor zeros,
respectively. We do not count them as nonstandard constraints
because a light neutrino mass matrix with more than two cofactor
zeros has already been shown to be excluded by experimental data.

\subsection{Class 4}
Our Class 4 constraints are equivalent to $M_{\alpha\alpha} = 0$,
$M_{\beta\beta}(M^{-1})_{\beta\beta}=1$ and $(M^{-1})_{\gamma\gamma}
\neq 0$ obtained in Ref.~\cite{Lavoura:2015wwa} as Eq.~(51). We demonstrate this equivalency below.

Since $M_{\alpha\alpha} = 0$,
\begin{align}
\det M=2 M_{\alpha\beta} M_{\beta\gamma} M_{\alpha\gamma}
- (M_{\alpha\gamma})^2M_{\beta\beta}
-(M_{\alpha\beta}^2)^2M_{\gamma\gamma}\,.
\end{align}
Plugging
$M_{\beta\beta}M_{\alpha\gamma}=2M_{\alpha\beta}M_{\beta\gamma}$ into
the above equation, we get
\begin{align}
\det M=-(M_{\alpha\beta}^2)^2M_{\gamma\gamma}\,.
\end{align}
Also, since $M_{\alpha\alpha} = 0$, we have
$(M^{-1})_{\gamma\gamma}=-\frac{(M_{\alpha\beta})^2}{\det M}$, and
therefore
\begin{align}
M_{\gamma\gamma}(M^{-1})_{\gamma\gamma}=1\,.
\end{align}
Together with $M_{\alpha\alpha} = 0$ and $M_{\beta\beta} \neq 0$
(which is implicitly included in our constraints), we see Class 4
constraints are consistent with the conditions (52) in
Ref.~\cite{Lavoura:2015wwa}. The form of the constraint in Table~\ref{tb:constraints} has the
merit of being quadratic instead of cubic (as in
Ref.~\cite{Lavoura:2015wwa}), which allows for easier numerical analysis.

The constraints $M_{\alpha\alpha} = 0$ and $M_{\beta\beta}M_{\alpha\gamma}=2M_{\alpha\beta}M_{\beta\gamma}$ can be written as
\begin{equation}
X_1+\sigma X_2+\rho X_3=0\,,
\label{eq:linear}
\end{equation}
and
\begin{equation}
A_1+\sigma^2 A_2+\rho^2 A_3+\sigma A_{12}+\rho A_{13}+\sigma\rho A_{23}=0\,,
\label{eq:quadratic}
\end{equation}
where $\sigma=(m_2/m_1)e^{-i\phi_2}$, $\rho=(m_3/m_1)e^{-i\phi_3}$, $X_i=U_{\alpha i}^{*2}$, $A_i=U_{\beta i}^{*2}U_{\alpha i}^*U_{\gamma i}^*$ and $A_{ij}=2U_{\beta i}^*U_{\beta j}^*(U_{\alpha i}^*U_{\gamma j}^*+U_{\alpha j}^*U_{\gamma i}^*)-(U_{\beta i}^{*2}U_{\alpha j}^*U_{\gamma j}^*+U_{\beta j}^{*2}U_{\alpha i}^*U_{\gamma i}^*)$ for $i=1,2,3$. Since Eq.~(\ref{eq:quadratic}) is quadratic, it can be solved with Eq.~(\ref{eq:linear}). We get
\begin{align}
\rho &= \frac{A_{23} X_1 X_2 X_3 - A_{13} X_2^2 X_3 - 2 A_2 X_1 X_3^2 + A_{12} X_2 X_3^2 \pm X_2\sqrt{\Lambda}}
{2 X_3 [A_3 X_2^2 + X_3 (A_2 X_3-A_{23} X_2 )]}
\,,\label{eq:rho}\\
\sigma &= \frac{A_{23} X_1 X_3 + A_{13} X_2 X_3 - 2 A_3 X_1 X_2 -A_{12} X_3^2 \mp \sqrt{\Lambda}}
{2 [A_3 X_2^2 + X_3 (A_2 X_3-A_{23} X_2 )]}
\,,\label{eq:sigma}
\end{align}
where $\Lambda= [2 A_3 X_1 X_2 + X_3 ( A_{12} X_3-A_{23} X_1 - A_{13} X_2 )]^2-4 [A_3 X_1^2 + X_3 ( A_1 X_3-A_{13} X_1)] [A_3 X_2^2 + X_3 (A_2 X_3-A_{23} X_2 )]$. Taking the absolute values of $\sigma$ and $\rho$, we find $m_2/m_1$ and $m_3/m_1$. Then $m_1$ can be written as
\begin{align}
m_1 &= \sqrt{\frac{\delta m^2 }{|\sigma|^2 - 1}}=\sqrt{\frac{\frac{1}{2}\delta m^2 \pm \Delta m^2 } {|\rho|^2 - 1}}\,,
\label{eq:m1}
\end{align}
where the plus (minus) sign in Eq.~(\ref{eq:m1}) is for the NH (IH).

Our numerical analysis proceeds as follows. Since there are 5
independent parameters in the light mass matrix in this case, we take
them to be $\theta_{12}$, $\theta_{23}$, $\theta_{13}$, $\delta m^2
\equiv m_2^2-m_1^2$ and $\Delta m^2 \equiv
m_3^2-(m_1^2+m_2^2)/2$. First, we choose a set of the five oscillation
parameters within the $2\sigma$ range from a recent
global fit~\cite{Capozzi:2013csa}. Then we solve Eq.~(\ref{eq:m1}) by scanning over
$\delta$ from 0 to $360^\circ$. We sample
$10^5$ sets of oscillation parameters allowed at $2\sigma$; by keeping
the values of $m_1$ and $\delta$ that are allowed, we obtain the
allowed regions in $m_1-\delta$ ($m_3-\delta$) plane for the NH (IH).

The allowed cases in Class 4 are as follows: 

\begin{enumerate}

\item $M_{\mu\mu} = 0$ and $M_{\tau\tau}M_{e\mu}=2M_{\mu \tau}M_{e
  \tau}$. This case is allowed for both hierarchies. The allowed
  regions are shown in Fig.~\ref{fg:c4-mte}.

\begin{figure}
\centering
\begin{subfigure}[b]{0.4\textwidth}
\includegraphics[width=\textwidth]{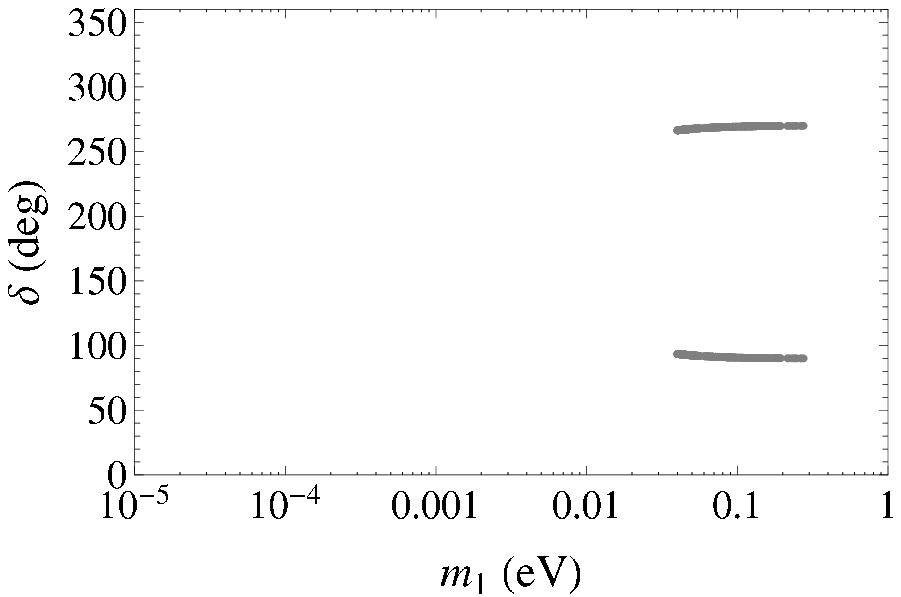}
\end{subfigure}
\quad\quad
\begin{subfigure}[b]{0.4\textwidth}
\includegraphics[width=\textwidth]{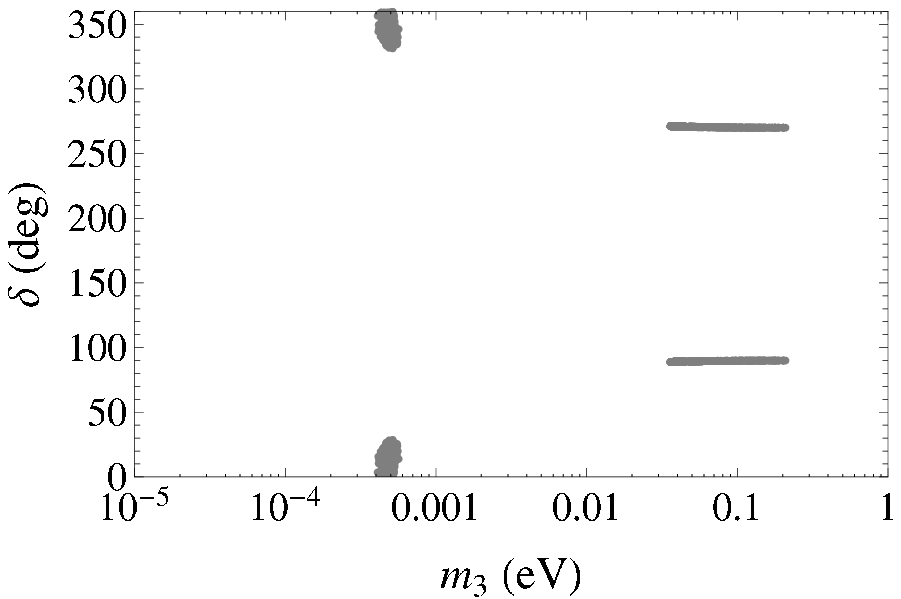}
\end{subfigure}
\caption{The $2\sigma$ allowed regions in the plane of the lightest mass and Dirac phase for the constraints, $M_{\mu\mu} = 0$ and $M_{\tau\tau}M_{e\mu}=2M_{\mu \tau}M_{e \tau}$. The left (right) panel is for the normal (inverted) hierarchy.}
\label{fg:c4-mte}
\end{figure}

\item $M_{\tau\tau} = 0$ and $M_{\mu\mu}M_{e\tau}=2M_{\mu \tau}M_{e
  \mu}$. This case is allowed for both hierarchies. The
  allowed regions are shown in Fig.~\ref{fg:c4-tme}.
  They are similar to Fig.~\ref{fg:c4-mte} with $\delta\rightarrow
  \delta+ 180^\circ$ for both hierarchies. The similarity is explained by the
  approximate $\mu-\tau$ symmetry of the experimental data.

\end{enumerate}


\begin{figure}
\centering
\begin{subfigure}[b]{0.4\textwidth}
\includegraphics[width=\textwidth]{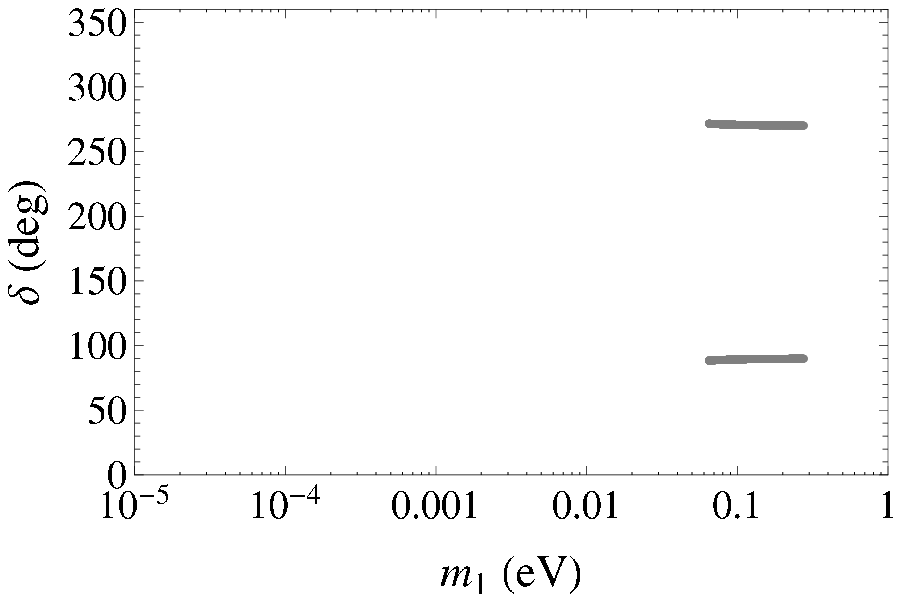}
\end{subfigure}
\quad\quad
\begin{subfigure}[b]{0.4\textwidth}
\includegraphics[width=\textwidth]{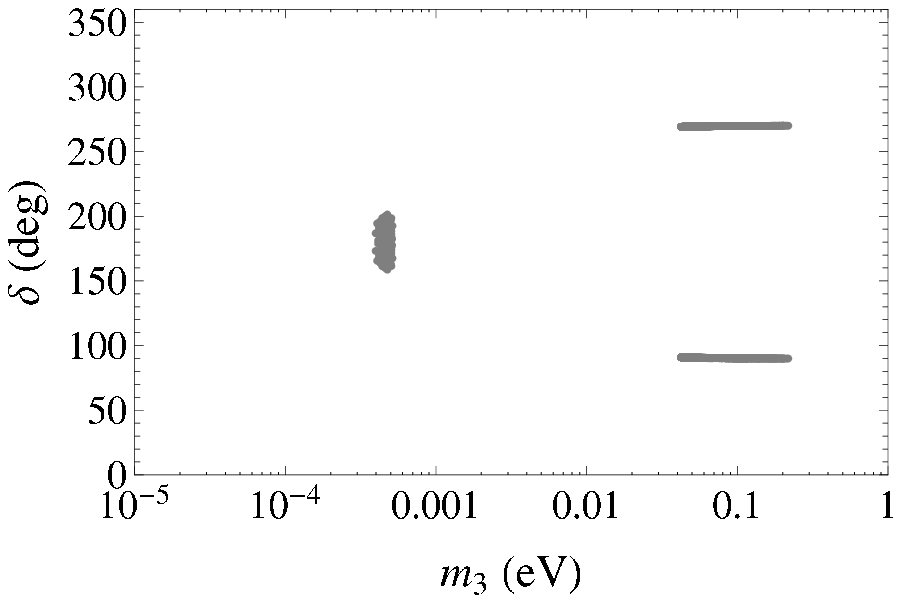}
\end{subfigure}
\caption{Same as Fig.~\ref{fg:c4-mte}, except for $M_{\tau\tau} = 0$ and $M_{\mu\mu}M_{e\tau}=2M_{\mu \tau}M_{e \mu}$. }
\label{fg:c4-tme}
\end{figure}

Note that these allowed regions are consistent with the results
obtained in Ref.~\cite{Lavoura:2015wwa}, in which 3$\sigma$ ranges of
each parameter are used as input.

\subsection{Class 5}

Similar to Class 4, Class 5 constraints are consistent with the
conditions $(M^{-1})_{\alpha\alpha} = 0$,
$M_{\beta\beta}(M^{-1})_{\beta\beta}=1$ and $M_{\gamma\gamma} \neq 0$
in Eq.~(53) of Ref.~\cite{Lavoura:2015wwa}. We find that it is easier to use
the quadratic form of Class 5 constraints for the numerical analysis.

Similar to Eqs.~(\ref{eq:linear}) and (\ref{eq:quadratic}), the two
constraints $C_{\alpha\alpha} = 0$ and
$C_{\beta\beta}C_{\beta\gamma}=2C_{\alpha\beta}C_{\alpha\gamma}$ can
be written as
\begin{equation}
X_1+ X_2/\sigma^*+ X_3/\rho^*=0\,,
\end{equation}
and
\begin{equation}
A_1+ A_2/(\sigma^*)^2+ A_3/(\rho^*)^2+ A_{12}/\sigma^*+ A_{13}/\rho^*+ A_{23}/(\sigma^*\rho^*)=0\,,
\end{equation}
where $\rho$, $\sigma$, $X_i$, $A_i$ and $A_{ij}$ are defined after Eq.~(\ref{eq:quadratic}). The analysis of this case follows that of Class 4 with the replacements, $\sigma\rightarrow 1/\sigma^*$ and $\rho\rightarrow
1/\rho^*$. The allowed cases in Class 5 are as follows:



\begin{enumerate}
\item $C_{ee} = 0$ and $C_{\mu\mu}C_{e\tau}=2C_{e\mu}C_{\mu\tau}$. This case is allowed for the IH only. The allowed regions are shown in the left panel of Fig.~\ref{fg:c5-IH}.

\begin{figure}
\centering
\begin{subfigure}[b]{0.4\textwidth}
\includegraphics[width=\textwidth]{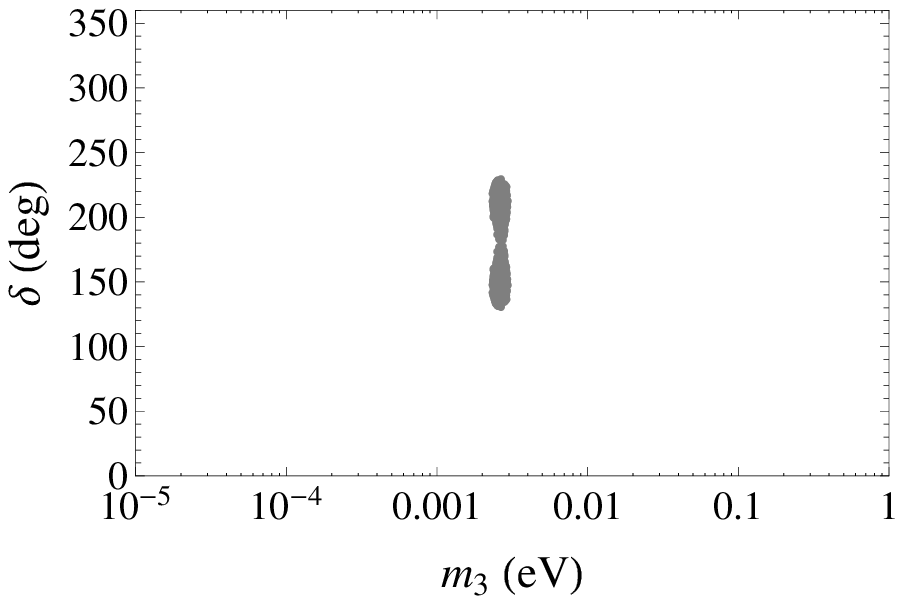}
\end{subfigure}
\quad\quad
\begin{subfigure}[b]{0.4\textwidth}
\includegraphics[width=\textwidth]{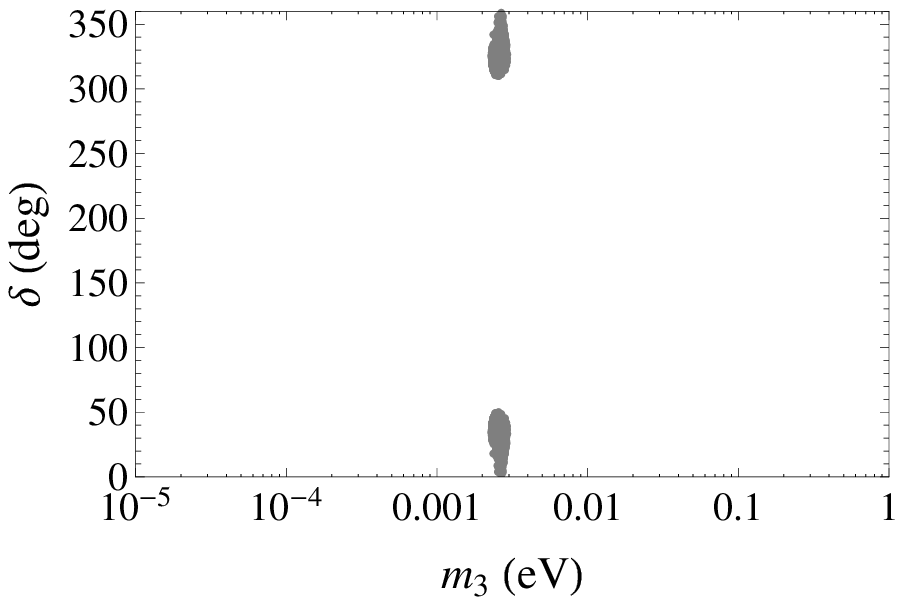}
\end{subfigure}
\caption{The $2\sigma$ allowed regions in the $(m_3,\delta)$ plane for the inverted hierarchy. The left (right) panel shows the results for the constraints, $C_{ee} = 0$ and $C_{\mu\mu}C_{e\tau}=2C_{e\mu}C_{\mu\tau}$ ($C_{\tau\tau}C_{e\mu}=2C_{e\tau}C_{\mu\tau}$). }
\label{fg:c5-IH}
\end{figure}


\item $C_{ee} = 0$ and $C_{\tau\tau}C_{e\mu}=2C_{e\tau}C_{\mu\tau}$. This case is allowed for the IH only. The allowed regions are shown in the right panel of Fig.~\ref{fg:c5-IH}, which is similar to the left panel with $\delta\rightarrow \delta+ 180^\circ$ because of the approximate $\mu-\tau$ symmetry. 


\item $C_{\mu\mu} = 0$ and $C_{\tau\tau}C_{e\mu}=2C_{\mu\tau}C_{e\tau}$. This case is allowed for both hierarchies. The allowed regions are shown in Fig.~\ref{fg:c5-mte}.

\begin{figure}
\centering
\begin{subfigure}[b]{0.4\textwidth}
\includegraphics[width=\textwidth]{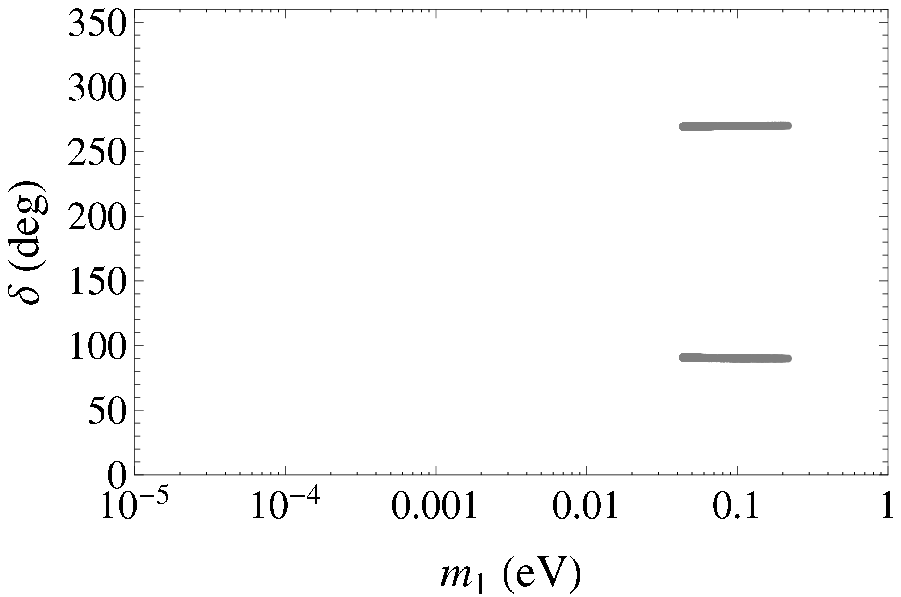}
\end{subfigure}
\quad\quad
\begin{subfigure}[b]{0.4\textwidth}
\includegraphics[width=\textwidth]{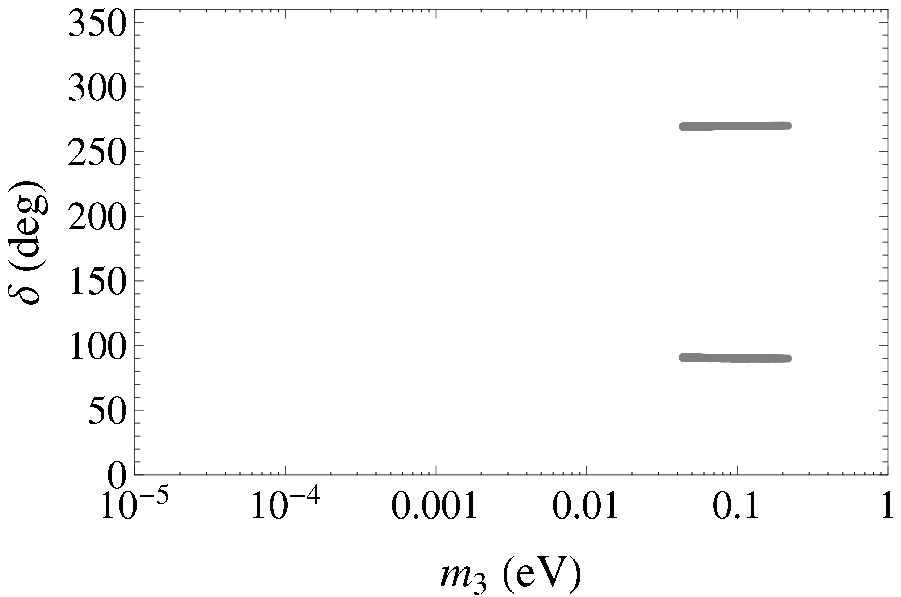}
\end{subfigure}
\caption{Same as Fig.~\ref{fg:c4-mte}, except for $C_{\mu\mu} = 0$ and $C_{\tau\tau}C_{e\mu}=2C_{\mu\tau}C_{e\tau}$. }
\label{fg:c5-mte}
\end{figure}

\item $C_{\tau\tau} = 0$ and
  $C_{\mu\mu}C_{e\tau}=2C_{\mu\tau}C_{e\mu}$. This case is allowed for
  both hierarchies. The allowed regions are similar to
  Fig.~\ref{fg:c5-mte} with
  $\delta\rightarrow \delta+ 180^\circ$ for both hierarchies. The similarity can be
  explained by the approximate $\mu-\tau$ symmetry of the experimental
  data.

\end{enumerate}

Note that these allowed regions are consistent with the results
obtained in Ref.~\cite{Lavoura:2015wwa}. Also, the Class 4 constraints
are dual to the Class 5 constraints because they have the same functional
form with the roles of elements and cofactors reversed~\cite{Liao:2013aka}. From
Ref.~\cite{Liao:2013aka}, we know that the allowed regions of Class 4
and 5 should be similar for opposite mass hierarchies when the
lightest mass is larger than about 20 meV. This can be seen from
the left panel of Fig.~\ref{fg:c4-mte} and the right panel of Fig.~\ref{fg:c5-mte}, and from
the right panel of Fig.~\ref{fg:c4-mte} and the left panel of Fig.~\ref{fg:c5-mte}.

\subsection{Class 6}
Since $\det M\neq 0$, the constraint
$M_{\alpha\alpha}C_{\alpha\alpha}=\det M$ is equivalent to
$M_{\alpha\alpha}(M^{-1})_{\alpha\alpha}=1$, which can be written as
\begin{align}
(m_1U_{\alpha 1}^{*2}+m_2U_{\alpha 2}^{*2}e^{-i\phi_2}+m_3U_{\alpha 3}^{*2}e^{-i\phi_3})(m_1^{-1}U_{\alpha 1}^2+m_2^{-1}U_{\alpha 2}^2e^{i\phi_2} +m_3^{-1}U_{\alpha 3}^2e^{i\phi_3})=1\,.
\label{eq:4+3}
\end{align}
Defining $U_{\alpha j}=|U_{\alpha j}|e^{i\beta j}$ and
$\phi'_j=\phi_j+2\beta_j-2\beta_1\,,$ we absorb the phases of the $U$'s
into $\phi_j$'s and write Eq.~(\ref{eq:4+3}) as
\begin{align}
(m_1|U_{\alpha 1}|^{2}+m_2|U_{\alpha 2}|^{2}e^{-i\phi'_2}+m_3|U_{\alpha 3}|^{2}e^{-i\phi'_3})
(m_1^{-1}|U_{\alpha 1}|^2+m_2^{-1}|U_{\alpha 2}|^2e^{i\phi' 2} +m_3^{-1}|U_{\alpha 3}|^2e^{i\phi' 3})=1\,.
\end{align}
We now expand the above equation into real and imaginary parts, and after some simplification we get
\begin{align}
&|U_{\alpha 1}|^4 + |U_{\alpha 2}|^4 + |U_{\alpha 3}|^4 +|U_{\alpha 2}|^2|U_{\alpha 3}|^2\left(\frac{m_2}{m_3}+\frac{m_3}{m_2}\right)(\cos\phi'_2\cos\phi'_3+\sin\phi'_2\sin\phi'_3)\nonumber\\
&+ |U_{\alpha 1}|^2|U_{\alpha 2}|^2\left(\frac{m_1}{m_2}+\frac{m_2}{m_1}\right)\cos\phi'_2 + |U_{\alpha 1}|^2|U_{\alpha 3}|^2\left(\frac{m_1}{m_3}+\frac{m_3}{m_1}\right)\cos\phi'_3=1\,,
\label{eq:2+5plug}
\end{align}
for the real part and
\begin{align}
&|U_{\alpha 1}|^2|U_{\alpha 2}|^2\left(\frac{m_1}{m_2}-\frac{m_2}{m_1}\right)\sin\phi'_2 + |U_{\alpha 1}|^2|U_{\alpha 3}|^2\left(\frac{m_1}{m_3}-\frac{m_3}{m_1}\right)\sin\phi'_3 \nonumber\\
& + |U_{\alpha 2}|^2|U_{\alpha 3}|^2\left(\frac{m_3}{m_2}-\frac{m_2}{m_3}\right)(\cos\phi'_3\sin\phi'_2-\sin\phi'_3\cos\phi'_2)=0\,,
\end{align}
for the imaginary part. The imaginary part can be put into the form
\begin{align}
D=A\cos\phi'_2+B\sin\phi'_2\,,
\label{eq:ABC}
\end{align}
where
\begin{align}
A&=-|U_{\alpha 2}|^2|U_{\alpha 3}|^2\left(\frac{m_3}{m_2}-\frac{m_2}{m_3}\right)\sin\phi'_3\,, \nonumber \\
B&=|U_{\alpha 2}|^2|U_{\alpha 3}|^2\left(\frac{m_3}{m_2}-\frac{m_2}{m_3}\right)\cos\phi'_3+|U_{\alpha 1}|^2|U_{\alpha 2}|^2\left(\frac{m_1}{m_2}-\frac{m_2}{m_1}\right)\,, \nonumber \\
D&=-|U_{\alpha 1}|^2|U_{\alpha 3}|^2\left(\frac{m_1}{m_3}-\frac{m_3}{m_1}\right)\sin\phi'_3\,. \nonumber
\end{align}
Equation~(\ref{eq:ABC}) has the solution (as long as $A^2+B^2\geq D^2$),
\begin{align}
\sin\phi'_2&=\frac{BD\pm A\sqrt{A^2+B^2-D^2}}{A^2+B^2}\,,\\
\cos\phi'_2&=\frac{AD\mp B\sqrt{A^2+B^2-D^2}}{A^2+B^2}\,.
\end{align}

For a given $\phi'_3$ we solve for $\phi'_2$ and plug that into Eq.~(\ref{eq:2+5plug}). Then for a fixed set of $\delta$, $m_1$ ($m_3$) for the NH (IH), and a given set of oscillation parameters, we scan $\phi'_3$ to see where (if) Eq.~(\ref{eq:2+5plug}) is satisfied. We also ignore those values of $\phi'_3$ that have no real solution for $\phi'_2$, i.e., $A^2+B^2<D^2$. If a solution is found, that point in $m_1-\delta$ space is allowed for the corresponding set of oscillation parameters. With the value of $\phi'_3$ that gives the solution, $\phi_2$ and $\phi_3$ are also obtained.

\begin{figure}
\centering
\begin{subfigure}[b]{0.4\textwidth}
\includegraphics[width=\textwidth]{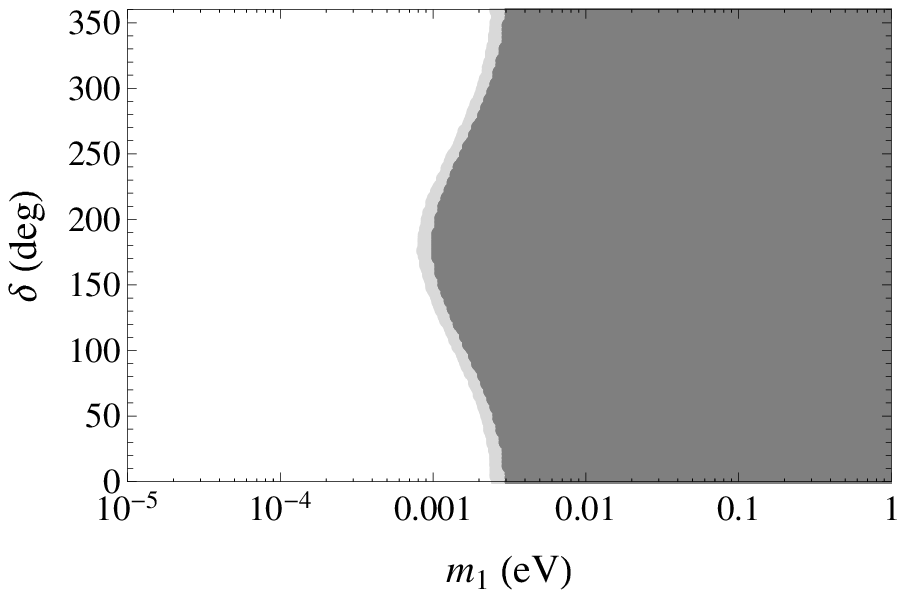}
\end{subfigure}
\quad\quad
\begin{subfigure}[b]{0.4\textwidth}
\includegraphics[width=\textwidth]{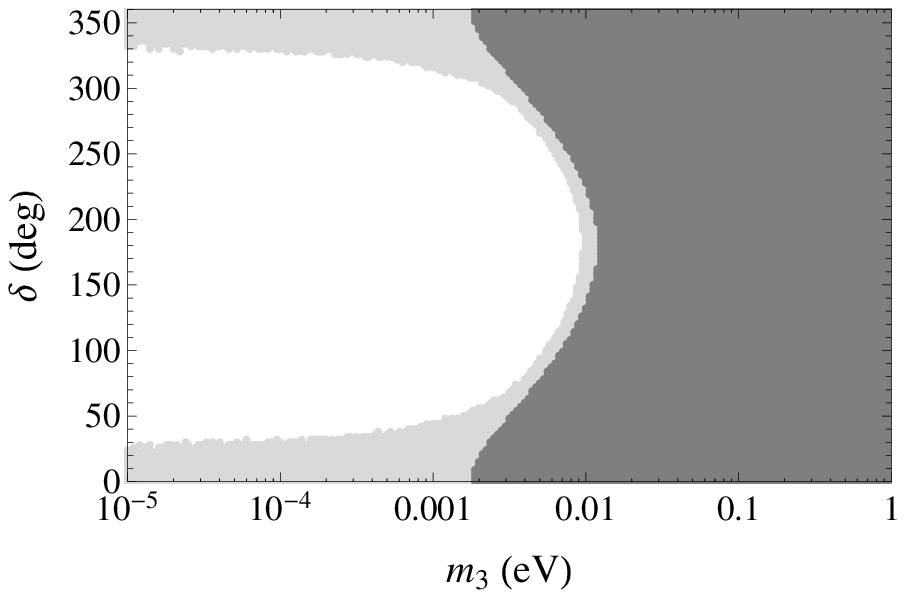}
\end{subfigure}
\caption{The allowed regions for the constraint $M_{\mu\mu}C_{\mu\mu}=\det M$. The left (right) panel is for the normal (inverted) hierarchy. The dark shaded regions correspond to the best-fit oscillation parameters, while the light
shaded regions are allowed at $2\sigma$.}
\label{fg:c6-m}
\end{figure}

\begin{figure}
\centering
\begin{subfigure}[b]{0.4\textwidth}
\includegraphics[width=\textwidth]{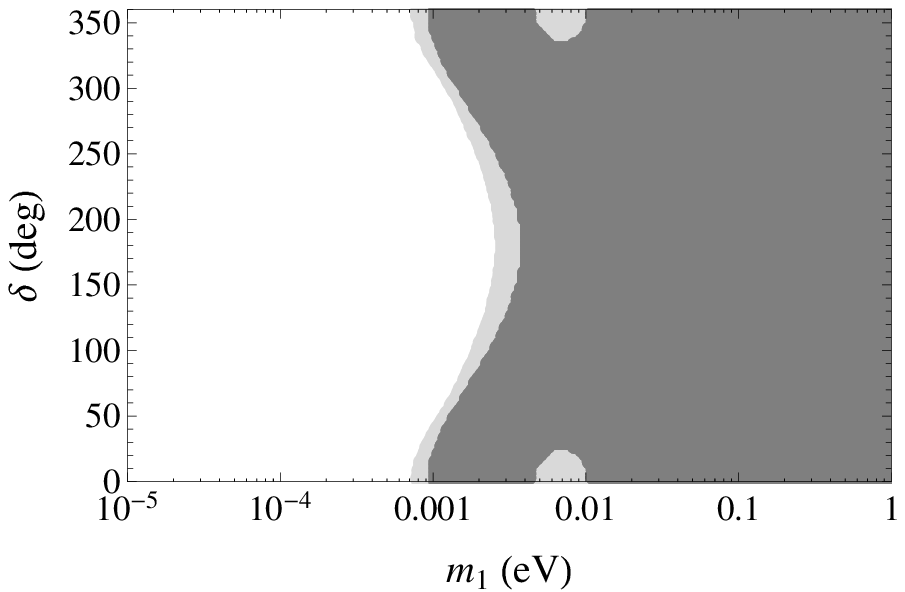}
\end{subfigure}
\quad\quad
\begin{subfigure}[b]{0.4\textwidth}
\includegraphics[width=\textwidth]{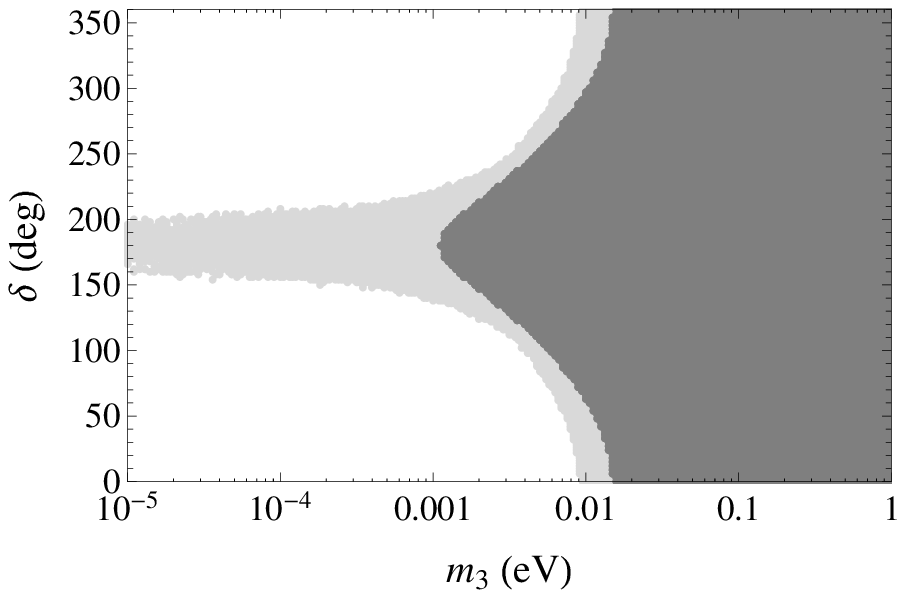}
\end{subfigure}
\caption{Same as Fig.~\ref{fg:c6-m}, except for $M_{\tau\tau}C_{\tau\tau}=\det M$.}
\label{fg:c6-t}
\end{figure}

The allowed cases in Class 6 are as follows: 

\begin{enumerate}

\item $M_{ee}C_{ee}=\det M$. This case is allowed for both hierarchies. There is no constraint on $\delta$ in this case. For the best-fit oscillation parameters, the allowed range of the lightest mass is $m_1>0.0024$ eV for the NH and $m_3>0.00050$ eV for the IH. The allowed range at $2\sigma$ is $m_1>0.0020$ eV for the NH and $m_3>0.00037$ eV for the IH.

\item $M_{\mu\mu}C_{\mu\mu}=\det M$. This case is allowed for both hierarchies, and the allowed regions are shown in Fig.~\ref{fg:c6-m}. 

\item $M_{\tau\tau}C_{\tau\tau}=\det M$. This case is allowed for both hierarchies, and the allowed regions are shown in Fig.~\ref{fg:c6-t}. The allowed regions are similar to Fig.~\ref{fg:c6-m} with $\delta\rightarrow \delta+ 180^\circ$ for both hierarchies because of the approximate $\mu-\tau$ symmetry of the data.

\end{enumerate}

\subsection{Class 7}
With $x=e^{-i\phi_2}$ and $y=e^{-i\phi_3}$, the constraint $M_{\alpha\alpha}^2C_{\alpha\alpha}=4M_{\alpha\beta}M_{\alpha\gamma}C_{\beta\gamma}$ can be written as
\begin{equation}
ax^3+bx^2+cx+d=0\,,
\label{eq:cubic}
\end{equation}
where
\begin{align}
a=\frac{m_2^2 U_{\alpha 2}^*{}^2}{m_1m_3}\left[(m_1U_{\alpha 3}^2+m_3 U_{\alpha 1}^2 y)U_{\alpha 2}^*{}^2-4(m_1U_{\beta 3}U_{\gamma 3}+m_3 U_{\beta 1}U_{\gamma 1} y)U_{\beta 2}^*U_{\gamma 2}^*\right]\,,
\end{align}
\begin{align}
b&=\frac{m_2U_{\alpha 2}^*}{m_1 m_3}\left\lbrace 2 m_1(m_1U_{\alpha 3}^2+m_3 U_{\alpha 1}^2y)U_{\alpha 1}^*{}^2U_{\alpha 2}^*\right. \nonumber \\
 &\left.-4 m_1(m_1U_{\beta 3}U_{\gamma 3}+m_3 U_{\beta 1}U_{\gamma 1}y)U_{\alpha 1}^*(U_{\beta 2}^*U_{\gamma 1}^*+U_{\beta 1}^*U_{\gamma 2}^*)\right. \nonumber \\
&\left.+ m_3 y\left[m_1 U_{\alpha 2}^2 U_{\alpha 2}^*{}^3+2 U_{\alpha 2}^*(m_1|U_{\alpha 3}|^4+m_3 y U_{\alpha 1}^2 U_{\alpha 3}^*{}^2-2 m_1 U_{\beta 2} U_{\gamma 2} U_{\beta 2}^* U_{\gamma 2}^* )\right.\right. \nonumber \\
&\left.\left.-4(m_1 U_{\beta 3}U_{\gamma 3}+m_3y U_{\beta 1}U_{\gamma 1})U_{\alpha 3}^*(U_{\beta 3}^*U_{\gamma 2}^*+U_{\beta 2}^*U_{\gamma 3}^*)\right] \right\rbrace\,,
\end{align}

\begin{align}
c&=\frac{1}{m_1 m_3}\left\lbrace m_1^2(m_1U_{\alpha 3}^2+m_3 U_{\alpha 1}^2y)U_{\alpha 1}^*{}^4 \right.\nonumber \\
&\left.+ 2m_1 U_{\alpha 1}^*{}^2\left[m_3y(m_1|U_{\alpha 2}|^4+m_1|U_{\alpha 3}|^4 + m_3 yU_{\alpha 1}^2 U_{\alpha 3}^*{}^2)\right.\right. \nonumber \\
&\left.\left.-2 m_1 (m_1 U_{\beta 3}U_{\gamma 3}+m_3y U_{\beta 1}U_{\gamma 1})U_{\beta 1}^*U_{\gamma 1}^*\right] \right. \nonumber \\
&-4m_1m_3y U_{\alpha 1}^*\left[m_1U_{\beta 2}U_{\gamma 2}U_{\alpha 2}^* (U_{\beta 2}^*U_{\gamma 1}^*+U_{\beta 1}^*U_{\gamma 2}^*)\right. \nonumber \\
& \left.+ (m_1 U_{\beta 3}U_{\gamma 3}+m_3y U_{\beta 1}U_{\gamma 1})U_{\alpha 3}^* (U_{\beta 3}^*U_{\gamma 1}^*+U_{\beta 1}^*U_{\gamma 3}^*) \right]  \nonumber \\
&\left.+m_3^2y^2U_{\alpha 3}^*\left[2 m_1 |U_{\alpha 2}|^4U_{\alpha 3}^*-4 (m_1 U_{\beta 3}U_{\gamma 3}+m_3y U_{\beta 1}U_{\gamma 1})U_{\alpha 3}^*U_{\beta 3}^*U_{\gamma 3}^* \right.\right. \nonumber \\
&\left.\left.+(m_1U_{\alpha 3}^2+m_3 U_{\alpha 1}^2y)(U_{\alpha 3}^*)^3- 4 m_1 U_{\beta 2}U_{\gamma 2}U_{\alpha 2}^*(U_{\beta 3}^*U_{\gamma 2}^*+U_{\beta 2}^*U_{\gamma 3}^*)\right] \right\rbrace\,,
\end{align}
\begin{align}
d&=\frac{y}{m_2}\left[ m_1^2 U_{\alpha 2}^2 U_{\alpha 1}^*{}^4+2 m_1 U_{\alpha 1}^*{}^2(m_3 U_{\alpha 2}^2 yU_{\alpha 3}^*{}^2-2 m_1 U_{\beta 2}U_{\gamma 2}U_{\beta 1}^*U_{\gamma 1}^*) \right.\nonumber \\
&\left. -4 m_1 m_3 U_{\beta 2} U_{\gamma 2} y U_{\alpha 1}^* U_{\alpha 3}^* (U_{\beta 3}^*U_{\gamma 1}^*+U_{\beta 1}^*U_{\gamma 3}^*)+ m_3^2 y^2 U_{\alpha 3}^*{}^2 (U_{\alpha 2}^2 U_{\alpha 3}^*{}^2-4 U_{\beta 2}U_{\gamma 2}U_{\beta 3}^*U_{\gamma 3}^*)\right]\,.
\end{align}

\begin{figure}
\centering
\includegraphics[width=0.8\textwidth]{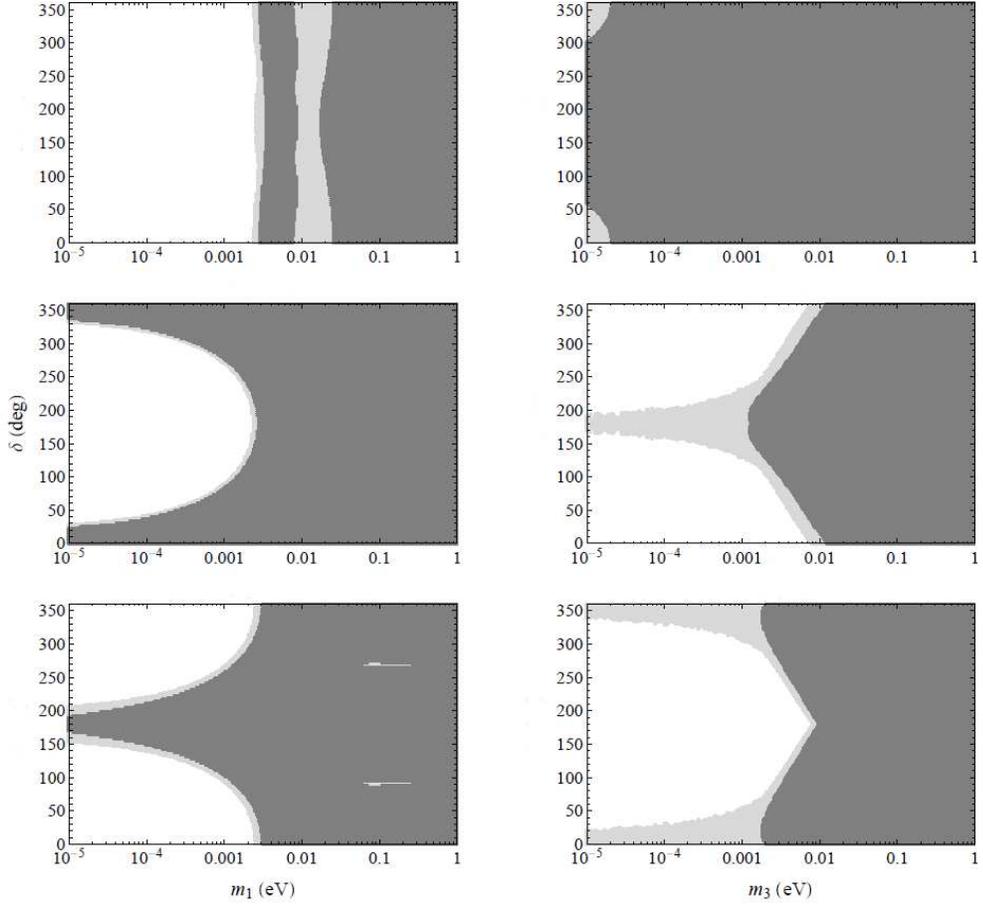}
%
%
\caption{The first, second and third rows show the allowed regions for the constraints, $M_{ee}^2C_{ee}=4M_{e\mu}M_{e\tau}C_{\mu\tau}$, $M_{\mu\mu}^2C_{\mu\mu}=4M_{e\mu}M_{\mu\tau}C_{e\tau}$ and $M_{\tau\tau}^2C_{\tau\tau}=4M_{e\tau}M_{\mu\tau}C_{e\mu}$, respectively. The dark shaded regions correspond to the best-fit oscillation parameters, while the light
shaded regions are allowed at $2\sigma$. 
The left (right) panels are for the normal (inverted) hierarchy. }
\label{fg:c7}
\end{figure}

For a fixed set of $\delta$, $m_1$ ($m_3$) for the NH (IH), and a given set of oscillation parameters, we solve the cubic equation (Eq.~\ref{eq:cubic}) to find $x$ as a function of $y$, and then scan over $\phi_3$ from 0 to $2\pi$ to find where (if) $|x|=1$. If $|x|=1$ is found, there is a solution for $\phi_2$ and $\phi_3$. By keeping the values of $m_1$ ($m_3$) and $\delta$ that have a solution, we obtain the allowed regions in the $m_1-\delta$ ($m_3-\delta$) space for the NH (IH). All the cases in Class 7 are allowed for the both hierarchies, and the allowed regions are shown in Fig.~\ref{fg:c7}. The 2$\sigma$ allowed regions in the second row of Fig.~\ref{fg:c7} are similar to the third row of Fig.~\ref{fg:c7} with $\delta\rightarrow \delta+ 180^\circ$ for both hierarchies because of the approximate $\mu-\tau$ symmetry of the data.

%

\subsection{Class 8}
Since $\det M\neq 0$, the constraint $M_{\beta\beta}\det M =-M_{\alpha\beta}^2C_{\alpha\alpha}$ is equivalent to 
\begin{equation}
M_{\beta\beta} = -M_{\alpha\beta}^2(M^{-1})_{\alpha\alpha}\,.
\end{equation}
Similar to Class 7, defining $x=e^{-i\phi_2}$ and $y=e^{-i\phi_3}$, we can write the constraint as
\begin{equation}
a'x^3+b'x^2+c'x+d'=0\,,
\end{equation}
where
\begin{align}
a'=\frac{m_2^2 U_{\text{$\alpha $2}}^*{}^2 U_{\text{$\beta $2}}^*{}^2 \left(m_1
   U_{\text{$\alpha $3}}^2+m_3 y U_{\text{$\alpha $1}}^2\right)}{m_1 m_3}\,,
\end{align}

\begin{align}
b'=m_2 U_{\text{$\beta $2}}^* \left[ y U_{\text{$\beta $2}}^* 
   \left(\left|U_{\text{$\alpha $2}}\right|{}^4+1\right)+2 U_{\text{$\alpha $2}}^*
   \left(ym_1^{-1}U_{\alpha 1}^2+m_3^{-1}U_{\alpha 3}^2\right) \left(m_1
   U_{\text{$\alpha $1}}^* U_{\text{$\beta $1}}^*+m_3 y U_{\text{$\alpha $3}}^*
   U_{\text{$\beta $3}}^* \right)\right]\,,
\end{align}

\begin{align}
c'=&y\left(m_1 U_{\text{$\beta $1}}^*{}^2+m_3 y U_{\text{$\beta $3}}^*{}^2\right)+2y U_{\text{$\alpha $2}}^*U_{\text{$\beta $2}}^*U_{\text{$\alpha $2}}^2\left(m_1U_{\text{$\alpha $1}}^*U_{\text{$\beta $1}}^*+m_3 y U_{\text{$\alpha $3}}^*U_{\text{$\beta $3}}^*\right)\\\nonumber
&+\left(m_1U_{\text{$\alpha $1}}^*U_{\text{$\beta $1}}^*+m_3 y U_{\text{$\alpha $3}}^*U_{\text{$\beta $3}}^*\right)^2\left(ym_1^{-1}U_{\alpha 1}^2+m_3^{-1}U_{\alpha 3}^2\right)\,,
\end{align}


\begin{align}
d'=\frac{y U_{\text{$\alpha $2}}^2 \left(m_1 U_{\text{$\alpha $1}}^* U_{\text{$\beta
   $1}}^*+m_3 U_{\text{$\alpha $3}}^* U_{\text{$\beta $3}}^* y\right){}^2}{m_2}\,.
\end{align}

\begin{figure}
\centering
\includegraphics[width=0.8\textwidth]{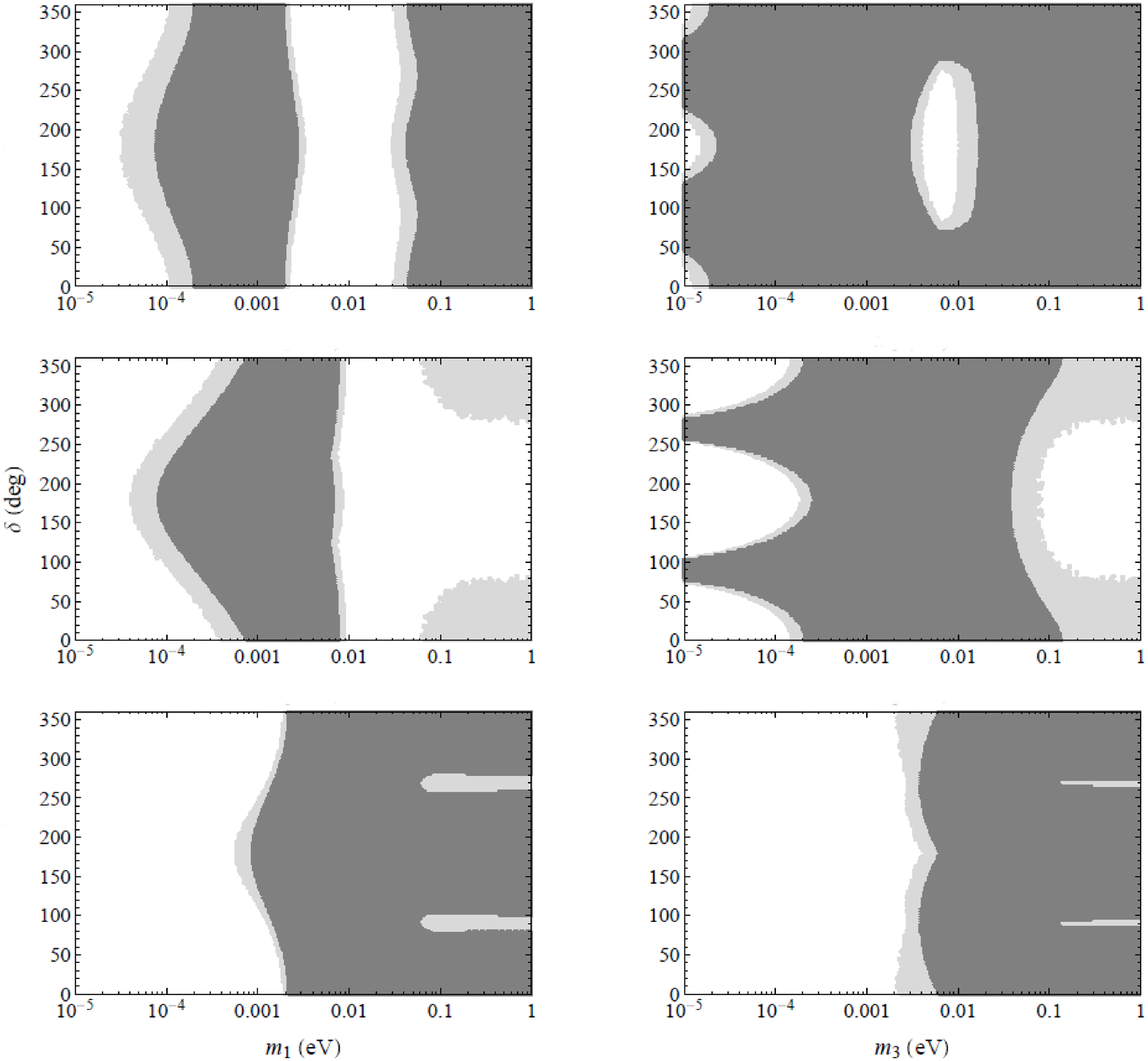}
%
%
\caption{Same as Fig.~\ref{fg:c7}, except that the first, second and third rows show the results for the constraints, $M_{\mu\mu}\det M =-M_{e\mu}^2C_{ee}$, $M_{ee}\det M =-M_{e\mu}^2C_{\mu\mu}$ and $M_{\tau\tau}\det M =-M_{\mu\tau}^2C_{\mu\mu}$, respectively.}
\label{fg:c8-a}
\end{figure}

\begin{figure}
\centering
\includegraphics[width=0.8\textwidth]{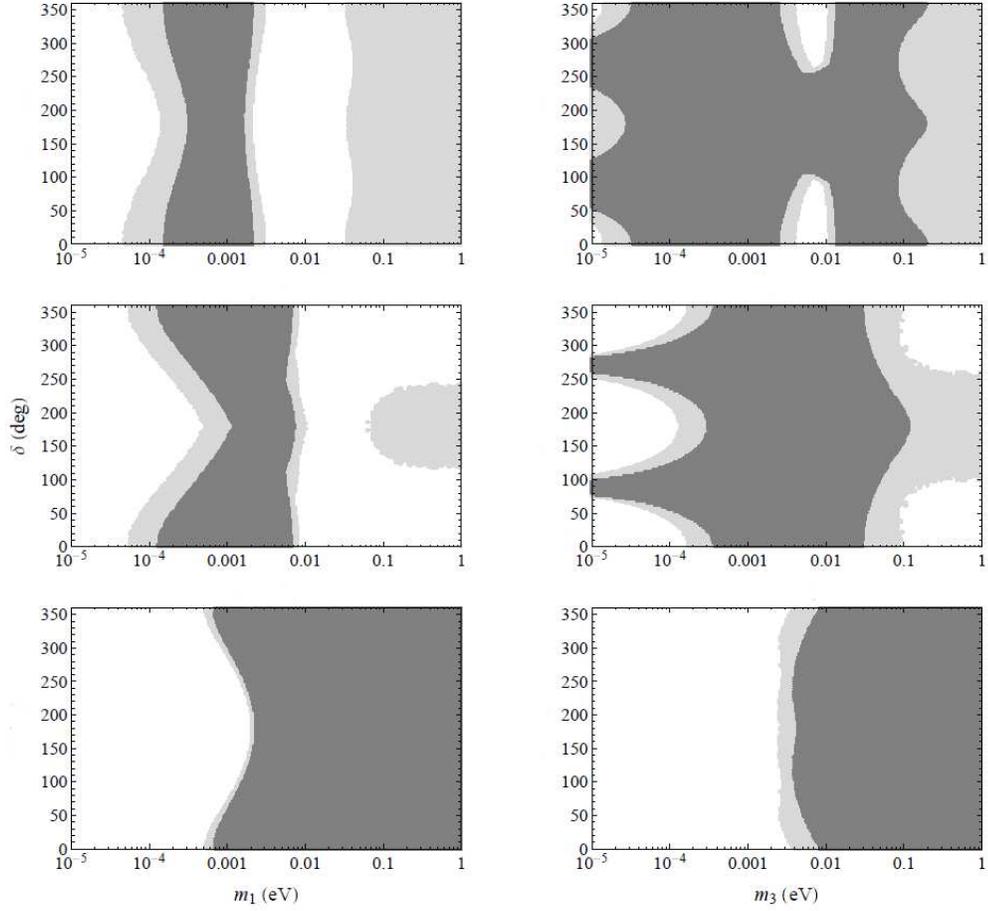}
%
%
\caption{Same as Fig.~\ref{fg:c7}, except that the first, second and third rows show the results for the constraints, $M_{\tau\tau}\det M =-M_{e\tau}^2C_{ee}$, $M_{ee}\det M =-M_{e\tau}^2C_{\tau\tau}$ and $M_{\mu\mu}\det M =-M_{\mu\tau}^2C_{\tau\tau}$, respectively.}
\label{fg:c8-b}
\end{figure}

The solutions for $\phi_2$ and $\phi_3$ then proceed as in Class 7. We find all the cases in Class 8 are allowed for the both hierarchies and the allowed regions are shown in Figs.~\ref{fg:c8-a} and~\ref{fg:c8-b}. Due to the approximate $\mu-\tau$ symmetry in the experimental data, the 2$\sigma$ allowed regions in each row of Fig.~\ref{fg:c8-a} are similar to those in the corresponding row of Fig.~\ref{fg:c8-b} with $\delta\rightarrow \delta+ 180^\circ$ for both hierarchies.

%

%

\section{Discussion and conclusion}
We carried out a systematic study of zeros in the Dirac and RH
Majorana neutrino mass matrices in the context of the Type I
seesaw mechanism. We derived complex constraints on the light neutrino mass
matrix for various textures in both the Dirac and RH
Majorana neutrino mass matrices. There are 9
nonstandard constraints besides the standard ones that can be
expressed in the form of one or two element/cofactor zeros alone. We
showed that both the standard and nonstandard constraints are stable
under one-loop RGE running from the lightest RH neutrino mass scale $M_1$
to the electroweak scale $M_Z$. In addition, we studied the
phenomenological implications for the nonstandard constraints, and
found that some cases for the normal or inverted hierarchy are excluded,
and for the rest we obtain the allowed regions for the lightest mass
and Dirac CP phase, which will be probed in the next generation of
neutrino experiments. We found 12 new models (Classes 6, 7 and 8) not
previously discussed in the literature which are allowed at $2\sigma$
for both the normal and inverted hierarchies.

Once the lightest neutrino mass and Dirac CP phase are determined, the rate for neutrinoless double beta decay ($0\nu\beta\beta$), which depends on the magnitude of the $\nu_e-\nu_e$
element of the neutrino mass matrix,
\begin{equation}
|M_{ee}| = |m_1c_{12}^2c_{13}^2 + m_2 e^{-i\phi_2}s_{12}^2c_{13}^2
+ m_3 e^{-i\phi_3}s_{13}^2e^{2i\delta}|\,,
\end{equation}
is fixed.

In Table~\ref{tab:Mee1}, we list the minimum and maximum values of
$|M_{ee}|$ at the $2\sigma$ level for Class 2, 3, 4 and 5 constraints. In
Table~\ref{tab:Mee2}, we list the minimum values for Class 6, 7, 8 and 9
constraints with the best-fit oscillation parameters and the $2\sigma$
lower bounds. The maximum values for Class 6, 7, 8 and 9 constraints are all
above 1000 meV. The most stringent experimental upper limit on the effective mass
$|M_{ee}|$ is $120-250$ meV at 90\%
C.L.~\cite{Gando:2012zm}. Some cases in Class 3, 4 and 5 could be ruled
out once the sensitivities to $|M_{ee}|$ reach about 50 meV for the
currently running $0\nu\beta\beta$
experiments~\cite{Rodejohann:2012xd}. Nevertheless, the minimum values of $|M_{ee}|$ for Class 6, 7, 8 and 9 constraints are all below $50$~meV and
can be completely probed only if the
sensitivity of future $0\nu\beta\beta$ experiments is significantly improved. Note that the minimum value of $|M_{ee}|$ is about 15 meV for the IH and 0 for the NH. The mass hierarchy will also be determined if $|M_{ee}|$ can be pushed to 15 meV in the future.

Recent global fits~\cite{Capozzi:2013csa, Forero:2014bxa, Gonzalez-Garcia:2014bfa} indicate a slight preference for a nonzero Dirac CP phase, with a preferred value $\delta \sim 270^\circ$.
Also, combined T2K and reactor analyses show a weak hint for the normal hierarchy~\cite{Abe:2015awa}.  Most viable cases of the nonstandard constraints allow $\delta=270^\circ$, except for two cases in Class 5: $C_{ee} = 0$, $C_{\mu\mu}C_{e\tau}=2C_{e\mu}C_{\mu\tau}$ and $C_{ee} = 0$, $C_{\tau\tau}C_{e\mu}=2C_{e\tau}C_{\mu\tau}$. These two cases may be ruled out if future measurements find $\delta$ to be close to $270^\circ$. If the hint for the normal hierarchy is confirmed, Classes 2 and 3, and the IH cases in the other classes will be ruled out.

\begin{table}
\begin{center}
\begin{tabular}{|c|c|c|c|c|}\hline
Class&Constraints &Hierarchy&Minimum&Maximum\\\hline
2&$M_{\mu\mu}=0$ and $\det M=0$ &IH&15&18\\\hline
2&$M_{\tau\tau}=0$ and $\det M=0$ &IH&15&17\\\hline
3&$M_{e\mu}=0$ and $\det M=0$ &IH&46&48\\\hline
3&$M_{e\tau}=0$ and $\det M=0$ &IH&46&48\\\hline
4&$M_{\mu\mu} = 0$ and $M_{\tau\tau}M_{e\mu}=2M_{\mu \tau}M_{e \tau}$ &NH&41&278\\\hline
4&$M_{\mu\mu} = 0$ and $M_{\tau\tau}M_{e\mu}=2M_{\mu \tau}M_{e \tau}$ &IH&16&214\\\hline
4&$M_{\tau\tau} = 0$ and $M_{\mu\mu}M_{e\tau}=2M_{\mu \tau}M_{e \mu}$ &NH&65&277\\\hline
4&$M_{\tau\tau} = 0$ and $M_{\mu\mu}M_{e\tau}=2M_{\mu \tau}M_{e \mu}$ &IH&15&223\\\hline
5&$C_{ee} = 0$ and $C_{\mu\mu}C_{e\tau}=2C_{e\mu}C_{\mu\tau}$ &IH&19&23\\\hline
5&$C_{ee} = 0$ and $C_{\tau\tau}C_{e\mu}=2C_{e\tau}C_{\mu\tau}$ &IH&19&24\\\hline
5&$C_{\mu\mu}=0$ and $C_{\tau\tau}C_{e\mu}=2C_{\mu\tau}C_{e\tau}$ &NH&64&278\\\hline
5&$C_{\mu\mu}=0$ and $C_{\tau\tau}C_{e\mu}=2C_{\mu\tau}C_{e\tau}$ &IH&65&222\\\hline
5&$C_{\tau\tau}=0$ and $C_{\mu\mu}C_{e\tau}=2C_{\mu\tau}C_{e\mu}$ &NH&39&268\\\hline
5&$C_{\tau\tau}=0$ and $C_{\mu\mu}C_{e\tau}=2C_{\mu\tau}C_{e\mu}$ &IH&61&219\\\hline
\end{tabular}
\end{center}
\caption{The minimum and maximum values of $|M_{ee}|$ (in meV) for Class 2, 3, 4 and 5 constraints at $2\sigma$.}
\label{tab:Mee1}
\end{table}

\begin{table}
\begin{center}
\begin{tabular}{|c|c|c|c|c|c|}\hline
\multirow{2}{*}{Class}&\multirow{2}{*}{Constraints}&\multicolumn{2}{c|}{Best-fit}&\multicolumn{2}{c|}{$2\sigma$ lower bound}\\\cline{3-6}
                      &&NH&IH&NH&IH\\\hline
6&$M_{ee}C_{ee}=\det M$ &3.2&17.9&2.6&15.1\\\hline
6&$M_{\mu\mu}C_{\mu\mu}=\det M$ & 0.2 & 17.9 & 0.0 & 14.9\\\hline
6&$M_{\tau\tau}C_{\tau\tau}=\det M$ &0.8&17.9&0.0&15.0\\\hline
7&$M_{ee}^2C_{ee}=4M_{e\mu}M_{e\tau}C_{\mu\tau}$ &0.0&41.4&0.0&39.6\\\hline
7&$M_{\mu\mu}^2C_{\mu\mu}=4M_{\mu\tau}M_{e\mu}C_{e\tau}$ &0.9&17.9&0.6&15.0\\\hline
7&$M_{\tau\tau}^2C_{\tau\tau}=4M_{e\tau}M_{\mu\tau}C_{e\mu}$ &0.8&17.9&0.5&15.0\\\hline
8&$M_{\mu\mu}\det M =-M_{e\mu}^2C_{ee}$ &1.6&18.0&1.3&14.9\\\hline
8&$M_{\tau\tau}\det M =-M_{e\tau}^2C_{ee}$ &1.7&17.9&1.4&14.9\\\hline
8&$M_{ee}\det M =-M_{e\mu}^2C_{\mu\mu}$ &0.0&21.7&0.0&18.1\\\hline
8&$M_{\tau\tau}\det M =-M_{\mu\tau}^2C_{\mu\mu}$ &0.5&17.9&0.2&15.0\\\hline
8&$M_{ee}\det M =-M_{e\tau}^2C_{\tau\tau}$ &0.0&20.3&0.0&18.6\\\hline
8&$M_{\mu\mu}\det M =-M_{\mu\tau}^2C_{\tau\tau}$ &0.5&17.9&0.2&15.0\\\hline
9&$\det M = 0$ &1.4&17.9&1.2&14.8\\\hline
\end{tabular}
\end{center}
\caption{The minimum values of $|M_{ee}|$ (in meV) for Class 6, 7, 8 and 9 constraints for the best-fit oscillation parameters, and the $2\sigma$ lower bounds.}
\label{tab:Mee2}
\end{table}
%

{\bf Acknowledgements.}
This research was supported by the DOE under grant No. DE-SC0010504 and by the Kavli
Institute for Theoretical Physics under NSF grant No. PHY11-25915.

\appendix

\section{Textures for the nonstandard constraints}
\label{ap:structure}
We list all possible structures that lead to the nonstandard constraints for $N=N_R+N_D\geq 7$. Also we do not consider $N_D\geq 7$ because it has one row of zeros in $M_D$, which leads to three zeros in $M$, which is excluded by experimental data. Following Ref.~\cite{Liao:2013saa}, we define Class X constraints on $M$ or $C\equiv\det (M) M^{-1}$ as one zero on diagonal and one off-diagonal zero sharing column and row, Class Y constraints as one zero on diagonal and one off-diagonal zero not sharing column and row, and Class Z constraints as two zeros on diagonal.
 
\subsection{$\bm{N_R=4}$}
For $N_R=4$, there is only one structure for $M_R$,  
\begin{align}
M_R=\begin{bmatrix}
   \times & 0 & 0 \\
   0 & 0 & \times  \\
   0 & \times & 0
   \end{bmatrix}\,,
\end{align}
where the symbol $\times$ denotes a nonzero matrix element. All other structures of $M_R$ can be obtained by a permutation of the columns and rows of the above structure. If $M_D$ has more than 5 zeros, $M$ or $C$ would have at least three zeros, so here we only consider $N_D\leq 5$.

\begin{enumerate}
\item \textit{$\bm{N_D=5}$.} Of the 126 different structures of $M_D$ in this case, 120 lead to at least three zeros in $M$ or $C$, and only 6 structures lead to Class 1 nonstandard constraints. The 6 structures are:

\begin{align}
M_D=\begin{bmatrix}
      0 & \times & 0 \\
      0 & 0 & \times  \\
      0 & \times & \times
   \end{bmatrix}\,,\quad\quad  \begin{bmatrix}
         0 & 0 & \times \\
         0 & \times & 0 \\
         0 & \times & \times  
      \end{bmatrix}\,,
\end{align}
which lead to the constraints, $M_{ee}=M_{\mu\mu}=0$ and $\det M=0$;
\begin{align}
M_D=\begin{bmatrix}
      0 & \times & 0 \\
      0 & \times & \times  \\
      0 & 0 & \times
   \end{bmatrix}\,,\quad\quad  \begin{bmatrix}
         0 & 0 & \times \\
         0 & \times & \times \\
         0 & \times & 0  
      \end{bmatrix}\,,
\end{align}
which lead to the constraints, $M_{ee}=M_{\tau\tau}=0$ and $\det M=0$;
\begin{align}
M_D=\begin{bmatrix}
      0 & \times & \times \\
      0 & \times & 0  \\
      0 & 0 & \times
   \end{bmatrix}\,,\quad\quad  \begin{bmatrix}
         0 & \times & \times \\
         0 & 0 & \times \\
         0 & \times & 0  
      \end{bmatrix}\,,
\end{align}
which lead to the constraints, $M_{\mu\mu}=M_{\tau\tau}=0$ and $\det M=0$.

\item \textit{$\bm{N_D=4}$.} Of the 126 different structures of $M_D$ in this case, 39 lead to at least three zeros in $M$ or $C$, 3 lead to a block diagonal matrix for $M$, 36 lead to Class X constraints on $M$, 6 lead to Class Z constraints on $M$, 6 lead to Class Z constraints on $C$, and 18 lead to one zero in $M$ and one zero in $C$. In addition, there are 6 structures of $M_D$ that lead to Class 2 nonstandard constraints:
\begin{align}
M_D=\begin{bmatrix}
      0 & \times & 0 \\
      0 & \times & \times  \\
      0 & \times & \times
   \end{bmatrix}\,,\quad\quad \begin{bmatrix}
            0 & 0 & \times \\
            0 & \times & \times \\
            0 & \times & \times  
         \end{bmatrix}\,,
\end{align}
which lead to the constraints, $M_{ee}=0$ and $\det M=0$;
\begin{align}
M_D=\begin{bmatrix}
      0 & \times & \times \\
      0 & \times & 0  \\
      0 & \times & \times
   \end{bmatrix}\,,\quad\quad  \begin{bmatrix}
            0 & \times & \times \\
            0 & 0 & \times \\
            0 & \times & \times  
         \end{bmatrix}\,,
\end{align}
which lead to the constraints, $M_{\mu\mu}=0$ and $\det M=0$;
\begin{align}
M_D=\begin{bmatrix}
      0 & \times & \times \\
      0 & \times & \times  \\
      0 & \times & 0
   \end{bmatrix}\,,\quad\quad  \begin{bmatrix}
            0 & \times & \times \\
            0 & \times & \times \\
            0 & 0 & \times  
         \end{bmatrix}\,,
\end{align}
which lead to the constraints, $M_{\tau\tau}=0$ and $\det M=0$.

The remaining 12 structures lead to Class 4 nonstandard constraints:
\begin{align}
M_D=\begin{bmatrix}
      0 & \times & 0 \\
      0 & \times & \times  \\
      \times & 0 & \times
   \end{bmatrix}\,, \quad\quad 
\begin{bmatrix}
         0 & 0 & \times \\
         0 & \times & \times  \\
         \times & \times & 0
      \end{bmatrix}\,,
\end{align}
which lead to the constraints, $M_{ee}=0$ and $M_{\mu\mu}M_{e\tau}=2M_{e\mu}M_{\mu\tau}$;
\begin{align}
M_D=\begin{bmatrix}
      0 & \times & 0 \\
      \times & 0 & \times  \\
      0 & \times & \times
   \end{bmatrix}\,, \quad\quad 
\begin{bmatrix}
         0 & 0 & \times \\
         \times & \times & 0  \\
         0 & \times & \times
      \end{bmatrix}\,,
\end{align}
which lead to the constraints, $M_{ee}=0$ and $M_{\tau\tau}M_{e\mu}=2M_{e\tau}M_{\mu\tau}$;
\begin{align}
M_D=\begin{bmatrix}
      0 & \times & \times \\
      0 & \times & 0  \\
      \times & 0 & \times
   \end{bmatrix}\,, \quad\quad 
\begin{bmatrix}
         0 & \times & \times \\
         0 & 0 & \times  \\
         \times & \times & 0
      \end{bmatrix}\,,
\end{align}
which lead to the constraints, $M_{\mu\mu}=0$ and $M_{ee}M_{\mu\tau}=2M_{e\mu}M_{e\tau}$;
\begin{align}
M_D=\begin{bmatrix}
      \times & 0 & \times \\
      0 & \times & 0  \\
      0 & \times & \times
   \end{bmatrix}\,, \quad\quad 
\begin{bmatrix}
         \times & \times & 0 \\
         0 & 0 & \times  \\
         0 & \times & \times
      \end{bmatrix}\,,
\end{align}
which lead to the constraints, $M_{\mu\mu}=0$ and $M_{\tau\tau}M_{e\mu}=2M_{\mu\tau}M_{e\tau}$;
\begin{align}
M_D=\begin{bmatrix}
      \times & 0 & \times \\
      0 & \times & \times  \\
      0 & \times & 0
   \end{bmatrix}\,, \quad\quad 
\begin{bmatrix}
         \times & \times & 0 \\
         0 & \times & \times  \\
         0 & 0 & \times
      \end{bmatrix}\,,
\end{align}
which lead to the constraints, $M_{\tau\tau}=0$ and $M_{\mu\mu}M_{e\tau}=2M_{e\mu}M_{\mu\tau}$;
\begin{align}
M_D=\begin{bmatrix}
      0 & \times & \times \\
      \times & 0 & \times  \\
      0 & \times & 0
   \end{bmatrix}\,, \quad\quad 
\begin{bmatrix}
         0 & \times & \times \\
         \times & \times & 0  \\
         0 & 0 & \times
      \end{bmatrix}\,,
\end{align}
which lead to the constraints, $M_{\tau\tau}=0$ and $M_{ee}M_{\mu\tau}=2M_{e\mu}M_{e\tau}$.

\item \textit{$\bm{N_D=3}$.} Of the 84 different structures of $M_D$ in this case, 5 lead to at least three zeros in $M$ or $C$, 12 lead to Class X constraints on $M$, 24 lead to one diagonal zero in $M$, 6 lead to one off-diagonal zero in $M$ and 24 lead to one diagonal zero in $C$. In addition, there are 6 structures that lead to Class 6 nonstandard constraints:
\begin{align}
M_D=\begin{bmatrix}
      \times & \times & 0 \\
      0 & \times & \times  \\
      0 & \times & \times
   \end{bmatrix}\,, \quad\quad 
   \begin{bmatrix}
            \times & 0 & \times \\
            0 & \times & \times  \\
            0 & \times & \times
         \end{bmatrix}\,,
\end{align}
which lead to the constraint, $M_{ee}C_{ee}=\det M$;
\begin{align}
M_D=\begin{bmatrix}
      0 & \times & \times  \\
      \times & \times & 0 \\
      0 & \times & \times
   \end{bmatrix}\,, \quad\quad 
   \begin{bmatrix}
            0 & \times & \times  \\
            \times & 0 & \times \\
            0 & \times & \times
         \end{bmatrix}\,,
\end{align}
which lead to the constraint, $M_{\mu\mu}C_{\mu\mu}=\det M$;
\begin{align}
M_D=\begin{bmatrix}
      0 & \times & \times \\
      0 & \times & \times  \\
      \times & \times & 0
   \end{bmatrix}\,, \quad\quad 
   \begin{bmatrix}
            0 & \times & \times \\
            0 & \times & \times  \\
            \times & 0 & \times
         \end{bmatrix}\,,
\end{align}
which lead to the constraint, $M_{\tau\tau}C_{\tau\tau}=\det M$. 

Also, there are 6 structures lead to Class 7 nonstandard constraints:
\begin{align}
M_D=\begin{bmatrix}
   0 & \times & \times \\
      \times & \times & 0  \\
      \times & 0 & \times
   \end{bmatrix}\,,\quad\quad 
      \begin{bmatrix}
               0 & \times & \times \\
               \times & 0 & \times  \\
               \times & \times & 0
            \end{bmatrix}\,,
\end{align}
which lead to the constraint, $M_{ee}^2C_{ee}=4M_{e\mu}M_{e\tau}C_{\mu\tau}$;
\begin{align}
M_D=\begin{bmatrix}
    \times & \times & 0\\
      0 & \times & \times  \\
      \times & 0 & \times
   \end{bmatrix}\,,\quad\quad 
      \begin{bmatrix}
               \times & 0 & \times \\
               0 & \times & \times  \\
               \times & \times & 0
            \end{bmatrix}\,,
\end{align}
which lead to the constraint, $M_{\mu\mu}^2C_{\mu\mu}=4M_{e\mu}M_{\mu\tau}C_{e\tau}$;
\begin{align}
M_D=\begin{bmatrix}
    \times & \times & 0 \\
      \times & 0 & \times  \\
      0 & \times & \times
   \end{bmatrix}\,,\quad\quad 
      \begin{bmatrix}
               \times & 0 & \times \\
               \times & \times & 0  \\
               0 & \times & \times
            \end{bmatrix}\,,
\end{align}
which lead to the constraint, $M_{\tau\tau}^2C_{\tau\tau}=4M_{e\tau}M_{\mu\tau}C_{e\mu}$.

The remaining one structure leads to the Class 9 constraint, $\det M=0$:
\begin{align}
M_D=\begin{bmatrix}
    0 & \times & \times \\
      0 & \times & \times  \\
      0 & \times & \times
   \end{bmatrix}\,.
\end{align}

\end{enumerate}

\subsection{$\bm{N_R=3}$}
For $N_R=3$, there are four structures of $M_R$ that are not equivalent under permutation. They are  
\begin{align}
M_{R1}=\begin{bmatrix}
   \times & 0 & 0 \\
   0 & \times & \times  \\
   0 & \times & 0
   \end{bmatrix}\,,\quad 
M_{R2}=\begin{bmatrix}
   \times & 0 & \times \\
   0 & 0 & \times  \\
   \times & \times & 0
   \end{bmatrix}\,,\nonumber \\
M_{R3}=\begin{bmatrix}
   0 & \times & \times \\
   \times & 0 & \times  \\
   \times & \times & 0
   \end{bmatrix}\,,\quad 
M_{R4}=\begin{bmatrix}
      \times & 0 & 0 \\
      0 & \times & 0  \\
      0 & 0 & \times
      \end{bmatrix}\,.
\end{align}
All other structures of $M_R$ can be obtained by a permutation of the columns and rows of the above structure. If $M_D$ has more than 5 zeros, $M$ or $C$ would have at least three zeros for all three structures of $M_R$, so here we only consider $N_D\leq 5$.

\subsubsection{$\bm{M_{R1}}$}
\begin{enumerate}
\item \textit{$\bm{N_D=5}$.} Of the 126 different structures of $M_D$ in this case, 84 lead to at least three zeros in $M$ or $C$, 6 lead to a block diagonal matrix for $M$, 12 lead to Class X constraints on $M$, and 12 lead to one zero in $M$ and one zero in $C$. In addition, there are 6 structures that lead to Class 2 nonstandard constraints:
\begin{align}
M_D=\begin{bmatrix}
      0 & \times & 0 \\
      0 & \times & \times  \\
      0 & 0 & \times
   \end{bmatrix}\,,\quad\quad
   \begin{bmatrix}
         0 & \times & 0 \\
         0 & 0 & \times  \\
         0 & \times & \times
      \end{bmatrix}\,,
\end{align}
which lead to the constraints, $M_{ee}=0$ and $\det M=0$;
\begin{align}
M_D=\begin{bmatrix}
      0 & \times & \times \\
      0 & \times & 0  \\
      0 & 0 & \times
   \end{bmatrix}\,,\quad\quad
   \begin{bmatrix}
         0 & 0 & \times \\
         0 & \times & 0  \\
         0 & \times & \times
      \end{bmatrix}\,,
\end{align}
which lead to the constraints, $M_{\mu\mu}=0$ and $\det M=0$;
\begin{align}
M_D=\begin{bmatrix}
      0 & 0 & \times \\
      0 & \times & \times  \\
      0 & \times & 0
   \end{bmatrix}\,,\quad\quad
   \begin{bmatrix}
         0 & \times & \times \\
         0 & 0 & \times  \\
         0 & \times & 0
      \end{bmatrix}\,,
\end{align}
which lead to the constraints, $M_{\tau\tau}=0$ and $\det M=0$.

The remaining 6 structures lead to Class 3 nonstandard constraints:
\begin{align}
M_D=\begin{bmatrix}
      \times & 0 & 0 \\
      0 & 0 & \times  \\
      \times & 0 & \times
   \end{bmatrix}\,,\quad\quad
\begin{bmatrix}
      0 & 0 & \times \\
      \times & 0 & 0  \\
      \times & 0 & \times
   \end{bmatrix}\,,
\end{align}
which lead to the constraints, $M_{e\mu}=0$ and $\det M=0$;
\begin{align}
M_D=\begin{bmatrix}
      \times & 0 & 0 \\
      \times & 0 & \times  \\
      0 & 0 & \times
   \end{bmatrix}\,,\quad\quad
\begin{bmatrix}
      0 & 0 & \times \\
      \times & 0 & \times  \\
      \times & 0 & 0
   \end{bmatrix}\,,
\end{align}
which lead to the constraints, $M_{e\tau}=0$ and $\det M=0$;
\begin{align}
M_D=\begin{bmatrix}
      \times & 0 & \times \\
      \times & 0 & 0  \\
      0 & 0 & \times
   \end{bmatrix}\,,\quad\quad
\begin{bmatrix}
      \times & 0 & \times \\
      0 & 0 & \times  \\
      \times & 0 & 0
   \end{bmatrix}\,,
\end{align}
which lead to the constraints, $M_{\mu\tau}=0$ and $\det M=0$.

\item \textit{$\bm{N_D=4}$.} Of the 126 different structures of $M_D$ in this case, 33 lead to at least three zeros in $M$ or $C$, 3 lead to a block diagonal matrix for $M$, 18 lead to Class X constraints on $M$, and 6 lead to one zero in $M$ and one zero in $C$, 15 lead to one diagonal zero in $M$, 12 lead to one off-diagonal zero in $M$, 9 lead to one diagonal zero in $C$, and 6 lead to one off-diagonal zero in $C$. In addition, there are 3 structures that lead to Class 2 nonstandard constraints:
\begin{align}
M_D=\begin{bmatrix}
      0 & \times & 0 \\
      0 & \times & \times  \\
      0 & \times & \times
   \end{bmatrix}\,,
\end{align}
which leads to the constraints, $M_{ee}=0$ and $\det M=0$;
\begin{align}
M_D=\begin{bmatrix}
      0 & \times & \times \\
      0 & \times & 0  \\
      0 & \times & \times
   \end{bmatrix}\,,
\end{align}
which leads to the constraints, $M_{\mu\mu}=0$ and $\det M=0$;
\begin{align}
M_D=\begin{bmatrix}
      0 & \times & \times \\
      0 & \times & \times  \\
      0 & \times & 0
   \end{bmatrix}\,,
\end{align}
which leads to the constraints, $M_{\tau\tau}=0$ and $\det M=0$.

Also, there are 6 structures that lead to Class 6 nonstandard constraints:
\begin{align}
M_D=\begin{bmatrix}
      \times & \times & 0 \\
      0 & \times & \times  \\
      0 & 0 & \times
   \end{bmatrix}\,,\quad\quad
   \begin{bmatrix}
         \times & \times & 0 \\
         0 & 0 & \times  \\
         0 & \times & \times
      \end{bmatrix}\,,
\end{align}
which lead to the constraint, $M_{ee}C_{ee}=\det M$;
\begin{align}
M_D=\begin{bmatrix}
      0 & \times & \times \\
      \times & \times & 0  \\
      0 & 0 & \times
   \end{bmatrix}\,,\quad\quad
   \begin{bmatrix}
         0 & 0 & \times \\
         \times & \times & 0  \\
         0 & \times & \times
      \end{bmatrix}\,,
\end{align}
which lead to the constraint, $M_{\mu\mu}C_{\mu\mu}=\det M$;
\begin{align}
M_D=\begin{bmatrix}
      0 & \times & \times \\
      0 & 0 & \times  \\
      \times & \times & 0
   \end{bmatrix}\,,\quad\quad
   \begin{bmatrix}
         0 & 0 & \times \\
         0 & \times & \times  \\
         \times & \times & 0
      \end{bmatrix}\,,
\end{align}
which lead to the constraint, $M_{\tau\tau}C_{\tau\tau}=\det M$.

There are 6 structures lead to Class 8 nonstandard constraints:
\begin{align}
M_D=\begin{bmatrix}
     \times & \times & 0 \\
        0 & 0 & \times  \\
        \times & 0 & \times
   \end{bmatrix}\,,
\end{align}
which leads to the constraint, $M_{\mu\mu}\det M =-M_{e\mu}^2C_{ee}$;
\begin{align}
M_D=\begin{bmatrix}
     \times & \times & 0 \\
        \times & 0 & \times  \\
        0 & 0 & \times
   \end{bmatrix}\,,
\end{align}
which leads to the constraint, $M_{\tau\tau}\det M =-M_{e\tau}^2C_{ee}$;
\begin{align}
M_D=\begin{bmatrix}
     0 & 0 & \times \\
        \times & \times & 0  \\
        \times & 0 & \times
   \end{bmatrix}\,,
\end{align}
which leads to the constraint, $M_{ee}\det M =-M_{e\mu}^2C_{\mu\mu}$;
\begin{align}
M_D=\begin{bmatrix}
     \times & 0 & \times \\
        \times & \times & 0  \\
        0 & 0 & \times
   \end{bmatrix}\,,
\end{align}
which leads to the constraint, $M_{\tau\tau}\det M =-M_{\mu\tau}^2C_{\mu\mu}$;
\begin{align}
M_D=\begin{bmatrix}
     \times & 0 & \times \\
        0 & 0 & \times  \\
        \times & \times & 0
   \end{bmatrix}\,,
\end{align}
which leads to the constraint, $M_{\mu\mu}\det M =-M_{\mu\tau}^2C_{\tau\tau}$;
\begin{align}
M_D=\begin{bmatrix}
     0 & 0 & \times \\
        \times & 0 & \times  \\
        \times & \times & 0
   \end{bmatrix}\,,
\end{align}
which leads to the constraint, $M_{ee}\det M =-M_{e\tau}^2C_{\tau\tau}$.

The remaining 9 structures lead to the Class 9 constraint, $\det M=0$:
\begin{align}
M_D=\begin{bmatrix}
      \times & 0 & \times \\
      \times & 0 & \times  \\
      \times & 0 & 0
   \end{bmatrix}\,,\quad\quad
   \begin{bmatrix}
         \times & 0 & \times \\
         \times & 0 & 0  \\
         \times & 0 & \times
      \end{bmatrix}\,,\quad\quad
      \begin{bmatrix}
               \times & 0 & 0 \\
               \times & 0 & \times  \\
               \times & 0 & \times
            \end{bmatrix}\,,\\
\begin{bmatrix}
      \times & 0 & \times \\
      \times & 0 & \times  \\
      0 & 0 & \times
   \end{bmatrix}\,,\quad\quad
   \begin{bmatrix}
         \times & 0 & \times \\
         0 & 0 & \times  \\
         \times & 0 & \times
      \end{bmatrix}\,,\quad\quad
      \begin{bmatrix}
               0 & 0 & \times \\
               \times & 0 & \times  \\
               \times & 0 & \times
            \end{bmatrix}\,,\\
\begin{bmatrix}
      0 & \times & \times \\
      0 & \times & \times  \\
      0 & 0 & \times
   \end{bmatrix}\,,\quad\quad
   \begin{bmatrix}
         0 & \times & \times \\
         0 & 0 & \times  \\
         0 & \times & \times
      \end{bmatrix}\,,\quad\quad
      \begin{bmatrix}
               0 & 0 & \times \\
               0 & \times & \times  \\
               0 & \times & \times
            \end{bmatrix}\,.
\end{align}

\end{enumerate}

\subsubsection{$\bm{M_{R2}}$}
\begin{enumerate}
\item \textit{$\bm{N_D=5.}$} Of the 126 different structures of $M_D$ in this case, 90 lead to at least three zeros in $M$ or $C$, 12 lead to Class X constraints on $M$, 6 lead to Class Z constraints on $C$, and 6 lead to one zero in $M$ and one zero in $C$. In addition, there are 6 structures that lead to Class 2 nonstandard constraints:
\begin{align}
M_D=\begin{bmatrix}
      0 & 0 & \times \\
      0 & \times & 0  \\
      0 & \times & \times
   \end{bmatrix}\,,\quad\quad
   \begin{bmatrix}
         0 & 0 & \times \\
         0 & \times & \times  \\
         0 & \times & 0
      \end{bmatrix}\,,
\end{align}
which lead to the constraints, $M_{ee}=0$ and $\det M=0$;
\begin{align}
M_D=\begin{bmatrix}
      0 & \times & 0 \\
      0 & 0 & \times  \\
      0 & \times & \times
   \end{bmatrix}\,,\quad\quad
   \begin{bmatrix}
         0 & \times & \times \\
         0 & 0 & \times  \\
         0 & \times & 0
      \end{bmatrix}\,,
\end{align}
which lead to the constraints, $M_{\mu\mu}=0$ and $\det M=0$;
\begin{align}
M_D=\begin{bmatrix}
      0 & \times & 0 \\
      0 & \times & \times  \\
      0 & 0 & \times
   \end{bmatrix}\,,\quad\quad
   \begin{bmatrix}
         0 & \times & \times \\
         0 & \times & 0  \\
         0 & 0 & \times
      \end{bmatrix}\,,
\end{align}
which lead to the constraints, $M_{\tau\tau}=0$ and $\det M=0$.

The remaining 6 structures lead to Class 5 nonstandard constraints:
\begin{align}
M_D=\begin{bmatrix}
   0 & \times & \times \\
   0 & \times & 0  \\
   \times & 0 & 0
   \end{bmatrix}\,,
\end{align}
which leads to the constraints, $C_{ee}=0$ and $C_{\mu\mu}C_{e\tau}=2C_{e\mu}C_{\mu\tau}$;
\begin{align}
M_D=\begin{bmatrix}
   0 & \times & \times \\
   \times & 0 & 0  \\
   0 & \times & 0
   \end{bmatrix}\,,
\end{align}
which leads to the constraints, $C_{ee}=0$ and $C_{\tau\tau}C_{e\mu}=2C_{e\tau}C_{\mu\tau}$;
\begin{align}
M_D=\begin{bmatrix}
   0 & \times & 0 \\
   0 & \times & \times  \\
   \times & 0 & 0
   \end{bmatrix}\,,
\end{align}
which leads to the constraints, $C_{\mu\mu}=0$ and $C_{ee}C_{\mu\tau}=2C_{e\mu}C_{e\tau}$;
\begin{align}
M_D=\begin{bmatrix}
   \times & 0 & 0 \\
   0 & \times & \times  \\
   0 & \times & 0
   \end{bmatrix}\,,
\end{align}
which leads to the constraints, $C_{\mu\mu}=0$ and $C_{\tau\tau}C_{e\mu}=2C_{\mu\tau}C_{e\tau}$;
\begin{align}
M_D=\begin{bmatrix}
   \times & 0 & 0 \\
   0 & \times & 0  \\
   0 & \times & \times
   \end{bmatrix}\,,
\end{align}
which leads to the constraints, $C_{\tau\tau}=0$ and $C_{\mu\mu}C_{e\tau}=2C_{\mu\tau}C_{e\mu}$;
\begin{align}
M_D=\begin{bmatrix}
   0 & \times & 0 \\
   \times & 0 & 0  \\
   0 & \times & \times
   \end{bmatrix}\,,
\end{align}
which leads to the constraints, $C_{\tau\tau}=0$ and $C_{ee}C_{\mu\tau}=2C_{e\tau}C_{e\mu}$.

\item \textit{$\bm{N_D=4.}$} Of the 126 different structures of $M_D$ in this case, 39 lead to at least three zeros in $M$ or $C$, 6 lead to Class Z constraints in $C$, 18 lead to Class X constraints on $M$, and 3 lead to one zero in $M$ and one zero in $C$, 12 lead to one diagonal zero in $M$, and 33 lead to one diagonal zero in $C$. In addition, there are 3 structures that lead to Class 2 nonstandard constraints:
\begin{align}
M_D=\begin{bmatrix}
      0 & 0 & \times \\
      0 & \times & \times  \\
      0 & \times & \times
   \end{bmatrix}\,,
\end{align}
which leads to the constraints, $M_{ee}=0$ and $\det M=0$;
\begin{align}
M_D=\begin{bmatrix}
      0 & \times & \times \\
      0 & 0 & \times  \\
      0 & \times & \times
   \end{bmatrix}\,,
\end{align}
which leads to the constraints, $M_{\mu\mu}=0$ and $\det M=0$;
\begin{align}
M_D=\begin{bmatrix}
      0 & \times & \times \\
      0 & \times & \times  \\
      0 & 0 & \times
   \end{bmatrix}\,,
\end{align}
which leads to the constraints, $M_{\tau\tau}=0$ and $\det M=0$.

Also, there are 9 structures leads to Class 6 nonstandard constraints:
\begin{align}
M_D=\begin{bmatrix}
      \times & 0 & \times \\
      0 & \times & \times  \\
      0 & \times & 0
   \end{bmatrix}\,, \quad\quad
\begin{bmatrix}
      \times & 0 & \times \\
      0 & \times & 0  \\
      0 & \times & \times
   \end{bmatrix}\,,\quad \quad
   \begin{bmatrix}
         \times & 0 & 0 \\
         0 & \times & \times  \\
         0 & \times & \times
      \end{bmatrix}\,,
\end{align}
which lead to the constraint, $M_{ee}C_{ee}=\det M$;
\begin{align}
M_D=\begin{bmatrix}
      0 & \times & \times \\
      \times & 0 & \times  \\
      0 & \times & 0
   \end{bmatrix}\,, \quad\quad
\begin{bmatrix}
      0 & \times & 0 \\
      \times & 0 & \times  \\
      0 & \times & \times
   \end{bmatrix}\,,\quad \quad
   \begin{bmatrix}
         0 & \times & \times \\
         \times & 0 & 0  \\
         0 & \times & \times
      \end{bmatrix}\,,
\end{align}
which lead to the constraint, $M_{\mu\mu}C_{\mu\mu}=\det M$;
\begin{align}
M_D=\begin{bmatrix}
      0 & \times & \times \\
      0 & \times & 0  \\
      \times & 0 & \times
   \end{bmatrix}\,, \quad\quad
\begin{bmatrix}
      0 & \times & 0 \\
      0 & \times & \times  \\
      \times & 0 & \times
   \end{bmatrix}\,,\quad \quad
   \begin{bmatrix}
         0 & \times & \times \\
         0 & \times & \times  \\
         \times & 0 & 0
      \end{bmatrix}\,,
\end{align}
which lead to the constraint, $M_{\tau\tau}C_{\tau\tau}=\det M$.

The remaining 3 structures lead to the Class 9 constraint, $\det M=0$:
\begin{align}
M_D=\begin{bmatrix}
      0 & \times & \times \\
      0 & \times & \times  \\
      0 & \times & 0
   \end{bmatrix}\,, \quad\quad
\begin{bmatrix}
      0 & \times & \times \\
      0 & \times & 0  \\
      0 & \times & \times
   \end{bmatrix}\,,\quad \quad
   \begin{bmatrix}
         0 & \times & 0 \\
         0 & \times & \times  \\
         0 & \times & \times
      \end{bmatrix}\,.
\end{align}

\end{enumerate}

\subsubsection{$\bm{M_{R3}}$}
\begin{enumerate}
\item \textit{$\bm{N_D=5}$.} Of the 126 different structures of $M_D$ in this case, 90 lead to at least three zeros in $M$ or $C$ and 36 lead to Class Z constraints on $C$.

\item \textit{$\bm{N_D=4}$.} Of the 126 different structures of $M_D$ in this case, 45 lead to at least three zeros in $M$ or $C$, 18 lead to Class Z constraints on $C$, and 63 lead to one diagonal zero in $C$.

\end{enumerate}

\subsubsection{$\bm{M_{R4}}$}
\begin{enumerate}
\item \textit{$\bm{N_D=5}$.} Of the 126 different structures of $M_D$ in this case, 72 lead to at least three zeros in $M$ or $C$ and 36 lead to a block diagonal matrix for $M$. The remaining 18 structures lead to Class 3 nonstandard constraints:
\begin{align}
M_D=\begin{bmatrix}
      \times & 0 & 0 \\
      0 & \times & 0  \\
      \times & \times & 0
   \end{bmatrix}\,,\quad\quad
  \begin{bmatrix}
         \times & 0 & 0 \\
         0 & 0 & \times  \\
         \times & 0 & \times
      \end{bmatrix}\,,\quad\quad
     \begin{bmatrix}
              0 & \times & 0 \\
              0 & 0 & \times  \\
              0 & \times & \times
           \end{bmatrix}\,,\\
\begin{bmatrix}
      0 & \times & 0 \\
      \times & 0 & 0  \\
      \times & \times & 0
   \end{bmatrix}\,,\quad\quad
  \begin{bmatrix}
         0 & 0 & \times \\
         \times & 0 & 0  \\
         \times & 0 & \times
      \end{bmatrix}\,,\quad\quad
     \begin{bmatrix}
              0 & 0 & \times \\
              0 & \times & 0  \\
              0 & \times & \times
           \end{bmatrix}\,,
\end{align}
which lead to the constraints, $M_{e\mu}=0$ and $\det M=0$;
\begin{align}
M_D=\begin{bmatrix}
      \times & 0 & 0 \\
      \times & \times & 0  \\
      0 & \times & 0
   \end{bmatrix}\,,\quad\quad
  \begin{bmatrix}
         \times & 0 & 0 \\
         \times & 0 & \times  \\
         0 & 0 & \times
      \end{bmatrix}\,,\quad\quad
     \begin{bmatrix}
              0 & \times & 0 \\
              0 & \times & \times  \\
              0 & 0 & \times
           \end{bmatrix}\,,\\
\begin{bmatrix}
      0 & \times & 0 \\
      \times & \times & 0  \\
      \times & 0 & 0
   \end{bmatrix}\,,\quad\quad
  \begin{bmatrix}
         0 & 0 & \times \\
         \times & 0 & \times  \\
         \times & 0 & 0
      \end{bmatrix}\,,\quad\quad
     \begin{bmatrix}
              0 & 0 & \times \\
              0 & \times & \times  \\
              0 & \times & 0
           \end{bmatrix}\,,
\end{align}
which lead to the constraints, $M_{e\tau}=0$ and $\det M=0$;
\begin{align}
M_D=\begin{bmatrix}
      \times & \times & 0 \\
      0 & \times & 0  \\
      \times & 0 & 0
   \end{bmatrix}\,,\quad\quad
  \begin{bmatrix}
         \times & 0 & \times \\
         0 & 0 & \times  \\
         \times & 0 & 0
      \end{bmatrix}\,,\quad\quad
     \begin{bmatrix}
              0 & \times & \times \\
              0 & 0 & \times  \\
              0 & \times & 0
           \end{bmatrix}\,,\\
\begin{bmatrix}
      \times & \times & 0 \\
      \times & 0 & 0  \\
      0 & \times & 0
   \end{bmatrix}\,,\quad\quad
  \begin{bmatrix}
         \times & 0 & \times \\
         \times & 0 & 0  \\
         0 & 0 & \times
      \end{bmatrix}\,,\quad\quad
     \begin{bmatrix}
              0 & \times & \times \\
              0 & \times & 0  \\
              0 & 0 & \times
           \end{bmatrix}\,,
\end{align}
which lead to the constraints, $M_{\mu\tau}=0$ and $\det M=0$.

\item \textit{$\bm{N_D=4}$.} Of the 126 different structures of $M_D$ in this case, 27 lead to at least three zeros in $M$ or $C$, 9 lead to a block diagonal matrix for $M$, 54 lead to one off-diagonal zero in $M$, and 18 lead to one off-diagonal zero in $C$. The remaining 18 structures lead to the Class 9 constraint, $\det M=0$:
\begin{align}
M_D=\begin{bmatrix}
      \times & \times & 0 \\
      \times & \times & 0  \\
      \times & 0 & 0
   \end{bmatrix}\,,\quad\quad
  \begin{bmatrix}
         \times & \times & 0 \\
         \times & 0 & 0  \\
         \times & \times & 0
      \end{bmatrix}\,,\quad\quad
     \begin{bmatrix}
              \times & 0 & 0 \\
              \times & \times & 0  \\
              \times & \times & 0
           \end{bmatrix}\,,\\
\begin{bmatrix}
      \times & \times & 0 \\
      \times & \times & 0  \\
      0 & \times & 0
   \end{bmatrix}\,,\quad\quad
  \begin{bmatrix}
         \times & \times & 0 \\
         0 & \times & 0  \\
         \times & \times & 0
      \end{bmatrix}\,,\quad\quad
     \begin{bmatrix}
              0 & \times & 0 \\
              \times & \times & 0  \\
              \times & \times & 0
           \end{bmatrix}\,,
\\
\begin{bmatrix}
      \times & 0 & \times \\
      \times & 0 & \times  \\
      \times & 0 & 0
   \end{bmatrix}\,,\quad\quad
  \begin{bmatrix}
         \times & 0 & \times \\
         \times & 0 & 0  \\
         \times & 0 & \times
      \end{bmatrix}\,,\quad\quad
     \begin{bmatrix}
              \times & 0 & 0 \\
              \times & 0 & \times  \\
              \times & 0 & \times
           \end{bmatrix}\,,\\
\begin{bmatrix}
      \times & 0 & \times \\
      \times & 0 & \times   \\
      0 & 0 & \times 
   \end{bmatrix}\,,\quad\quad
  \begin{bmatrix}
         \times & 0 & \times  \\
         0 & 0 & \times   \\
         \times & 0 & \times 
      \end{bmatrix}\,,\quad\quad
     \begin{bmatrix}
              0 & 0 & \times  \\
              \times & 0 & \times   \\
              \times & 0 & \times 
           \end{bmatrix}\,,
\\
\begin{bmatrix}
      0 & \times &  \times \\
      0 & \times &  \times  \\
      0 & \times &  0
   \end{bmatrix}\,,\quad\quad
  \begin{bmatrix}
         0 & \times &  \times \\
         0 & \times &  0  \\
         0 & \times &  \times
      \end{bmatrix}\,,\quad\quad
     \begin{bmatrix}
              0 & \times &  0 \\
              0 & \times &  \times  \\
              0 & \times &  \times
           \end{bmatrix}\,,\\
\begin{bmatrix}
      0 & \times &  \times \\
      0 & \times &  \times   \\
      0 & 0 &  \times 
   \end{bmatrix}\,,\quad\quad
  \begin{bmatrix}
         0 & \times &  \times  \\
         0 & 0 & \times   \\
          0 & \times & \times 
      \end{bmatrix}\,,\quad\quad
     \begin{bmatrix}
              0 & 0 & \times  \\
              0 & \times &  \times   \\
              0 & \times &  \times 
           \end{bmatrix}\,.
\end{align}
Since $M_R$ is diagonal in this case, the constraints mainly come from four zeros in $M_D$, and they are consistent with the results discussed in Refs.~\cite{Liao:2013kix, Branco:2007nb}. 
\end{enumerate}

\subsection{$\bm{N_R=2}$}
If $M_R$ has two zeros, there are four structures of $M_R$ that are not equivalent under permutation. They are  
\begin{align}
M_{Ra}=\begin{bmatrix}
   0 & 0 & \times \\
   0 & \times & \times  \\
   \times & \times & \times
   \end{bmatrix}\,,\quad 
M_{Rb}=\begin{bmatrix}
   0 & \times & \times \\
   \times & \times & 0  \\
   \times & 0 & \times
   \end{bmatrix}\,,\nonumber \\
M_{Rc}=\begin{bmatrix}
   \times & 0 & 0 \\
   0 & \times & \times  \\
   0 & \times & \times
   \end{bmatrix}\,,\quad
M_{Rd}=\begin{bmatrix}
   \times & \times & \times \\
   \times & 0 & \times  \\
   \times & \times & 0
   \end{bmatrix}\,.
\end{align}
All other structures of $M_R$ can be obtained by a permutation of the columns and rows of the above structure. If $M_D$ have more than 6 zeros, $M$ or $C$ would have at least three zeros for all three structures of $M_R$, so here we only consider $N_D\leq 6$.

\subsubsection{$\bm{M_{Ra}}$}
\begin{enumerate}
\item \textit{$\bm{N_D=6}$.} Of the 84 different structures of $M_D$ in this case, 78 lead to at least three zeros in $M$ or $C$, and 6 lead to Class X constraints for $M$.

\item \textit{$\bm{N_D=5}$.} Of the 126 different structures of $M_D$ in this case, 78 lead to at least three zeros in $M$ or $C$, 18 lead to Class X constraints on $M$, 6 lead to one diagonal zero in $M$, and 6 lead to one diagonal zero in $C$. In addition, there are 6 structures that lead to Class 2 nonstandard constraints:
\begin{align}
M_D=\begin{bmatrix}
      0 & 0 & \times \\
      \times & 0 & \times  \\
      \times & 0 & 0
   \end{bmatrix}\,,\quad\quad
\begin{bmatrix}
      0 & 0 & \times \\
      \times & 0 & 0  \\
      \times & 0 & \times
   \end{bmatrix}\,,
\end{align}
which lead to the constraints, $M_{ee}=0$ and $\det M=0$;  
\begin{align}
M_D=\begin{bmatrix}
      \times & 0 & \times \\
      0 & 0 & \times  \\
      \times & 0 & 0
   \end{bmatrix}\,,\quad \quad
\begin{bmatrix}
      \times & 0 & 0 \\
      0 & 0 & \times  \\
      \times & 0 & \times
   \end{bmatrix}\,,
\end{align}
which lead to the constraints, $M_{\mu\mu}=0$ and $\det M=0$;
\begin{align}
M_D=\begin{bmatrix}
      \times & 0 & \times \\
      \times & 0 & 0  \\
      0 & 0 & \times
   \end{bmatrix}\,,\quad \quad
\begin{bmatrix}
      \times & 0 & 0 \\
      \times & 0 & \times  \\
      0 & 0 & \times
   \end{bmatrix}\,,
\end{align}
which lead to the constraints, $M_{\tau\tau}=0$ and $\det M=0$.

Also, there are 6 structures lead to Class 6 nonstandard constraints:
\begin{align}
M_D=\begin{bmatrix}
      0 & \times & 0 \\
      \times & 0 & \times  \\
      \times & 0 & 0
   \end{bmatrix}\,,\quad \quad
   \begin{bmatrix}
         0 & \times & 0 \\
         \times & 0 & 0  \\
         \times & 0 & \times
      \end{bmatrix}\,,
\end{align}
which lead to the constraint, $M_{ee}C_{ee}=\det M$;
\begin{align}
M_D=\begin{bmatrix}
      \times & 0 & \times \\
      0 & \times & 0  \\
      \times & 0 & 0
   \end{bmatrix}\,,\quad \quad
   \begin{bmatrix}
         \times & 0 & 0 \\
         0 & \times & 0  \\
         \times & 0 & \times
      \end{bmatrix}\,,
\end{align}
which lead to the constraint, $M_{\mu\mu}C_{\mu\mu}=\det M$;
\begin{align}
M_D=\begin{bmatrix}
      \times & 0 & 0 \\
      \times & 0 & \times  \\
      0 & \times & 0
   \end{bmatrix}\,,\quad \quad
   \begin{bmatrix}
         \times & 0 & \times \\
         \times & 0 & 0  \\
         0 & \times & 0
      \end{bmatrix}\,,
\end{align}
which lead to the constraint, $M_{\tau\tau}C_{\tau\tau}=\det M$.

The remaining 6 structures lead to the Class 9 constraint, $\det M=0$:
\begin{align}
M_D=\begin{bmatrix}
      \times & \times & 0 \\
      \times & 0 & 0  \\
      0 & \times & 0
   \end{bmatrix}\,,\quad \quad
   \begin{bmatrix}
         \times & \times & 0 \\
         0 & \times & 0  \\
         \times & 0 & 0
      \end{bmatrix}\,,\quad \quad
         \begin{bmatrix}
               \times & 0 & 0 \\
               \times & \times & 0  \\
               0 & \times & 0
            \end{bmatrix}\,,
\\
\begin{bmatrix}
      \times & 0 & 0 \\
      0 & \times & 0  \\
      \times & \times & 0
   \end{bmatrix}\,,\quad \quad
   \begin{bmatrix}
         0 & \times & 0 \\
         \times & \times & 0  \\
         \times & 0 & 0
      \end{bmatrix}\,,\quad \quad
         \begin{bmatrix}
               0 & \times & 0 \\
               \times & 0 & 0  \\
               \times & \times & 0
            \end{bmatrix}\,.
\end{align}

\end{enumerate}

\subsubsection{$\bm{M_{Rb}}$}
\begin{enumerate}
\item \textit{$\bm{N_D=6}$.} Of the 84 different structures of $M_D$ in this case, 78 lead to at least three zeros in $M$ or $C$, and 6 lead to Class Y constraints on $C$.

\item \textit{$\bm{N_D=5}$.} Of the 126 different structures of $M_D$ in this case, 78 lead to at least three zeros in $M$ or $C$, 24 lead to one diagonal zero in $C$, and 12 lead to one off-diagonal zero in $C$. The remaining 12 structures lead to the Class 9 constraint, $\det M=0$:
\begin{align}
M_D=\begin{bmatrix}
      \times & \times & 0 \\
      \times & 0 & 0  \\
      0 & \times & 0
   \end{bmatrix}\,,\quad \quad
   \begin{bmatrix}
         \times & \times & 0 \\
         0 & \times & 0  \\
         \times & 0 & 0
      \end{bmatrix}\,,\quad \quad
         \begin{bmatrix}
               \times & 0 & 0 \\
               \times & \times & 0  \\
               0 & \times & 0
            \end{bmatrix}\,,
\\
\begin{bmatrix}
      \times & 0 & 0 \\
      0 & \times & 0  \\
      \times & \times & 0
   \end{bmatrix}\,,\quad \quad
   \begin{bmatrix}
         0 & \times & 0 \\
         \times & \times & 0  \\
         \times & 0 & 0
      \end{bmatrix}\,,\quad \quad
         \begin{bmatrix}
               0 & \times & 0 \\
               \times & 0 & 0  \\
               \times & \times & 0
            \end{bmatrix}\,,
\\
\begin{bmatrix}
      \times  & 0 & \times\\
      \times & 0 & 0  \\
      0 & 0 & \times 
   \end{bmatrix}\,,\quad \quad
   \begin{bmatrix}
         \times  & 0 & \times\\
         0  & 0 & \times \\
         \times & 0 & 0
      \end{bmatrix}\,,\quad \quad
         \begin{bmatrix}
               \times & 0 & 0 \\
               \times  & 0 & \times \\
               0  & 0 & \times
            \end{bmatrix}\,,
\\
\begin{bmatrix}
      \times & 0 & 0 \\
      0  & 0 & \times \\
      \times  & 0 & \times
   \end{bmatrix}\,,\quad \quad
   \begin{bmatrix}
         0  & 0 & \times \\
         \times  & 0  & \times\\
         \times & 0 & 0
      \end{bmatrix}\,,\quad \quad
         \begin{bmatrix}
               0  & 0 & \times\\
               \times & 0 & 0  \\
               \times  & 0 & \times
            \end{bmatrix}\,.
\end{align}

\end{enumerate}

\subsubsection{$\bm{M_{Rc}}$}
\begin{enumerate}
\item \textit{$\bm{N_D=6}$.} Of the 84 different structures of $M_D$ in this case, 78 lead to at least three zeros in $M$ or $C$, and 6 lead to a block diagonal matrix for $M$.

\item \textit{$\bm{N_D=5}$.} Of the 126 different structures of $M_D$ in this case, 72 lead to at least three zeros in $M$ or $C$, 12 lead to a block diagonal matrix for $M$, 24 lead to one off-diagonal zero in $M$, and 12 lead to one off-diagonal zero in $C$. The remaining 6 structures lead to the Class 9 constraint, $\det M=0$:
\begin{align}
M_D=\begin{bmatrix}
      0 & \times & \times  \\
      0 & \times & 0   \\
      0 & 0 & \times 
   \end{bmatrix}\,,\quad \quad
   \begin{bmatrix}
         0 & \times & \times \\
         0 & 0 & \times   \\
         0 & \times & 0 
      \end{bmatrix}\,,\quad \quad
         \begin{bmatrix}
               0 & \times & 0  \\
               0 & \times & \times   \\
               0 & 0 & \times 
            \end{bmatrix}\,,
\\
\begin{bmatrix}
      0 & \times & 0 \\
      0 & 0 & \times  \\
      0 & \times & \times 
   \end{bmatrix}\,,\quad \quad
   \begin{bmatrix}
         0 & 0 & \times  \\
         0 & \times & \times  \\
         0 & \times & 0 
      \end{bmatrix}\,,\quad \quad
         \begin{bmatrix}
               0 & 0 & \times  \\
               0 & \times & 0   \\
               0 & \times & \times 
            \end{bmatrix}\,.
\end{align}

\end{enumerate}

\subsubsection{$\bm{M_{Rd}}$}
\begin{enumerate}
\item \textit{$\bm{N_D=6}$.} Of the 84 different structures of $M_D$ in this case, 78 lead to at least three zeros in $M$ or $C$, and 6 lead to  Class Z constraints on $M$.

\item \textit{$\bm{N_D=5}$.} Of the 126 different structures of $M_D$ in this case, 84 lead to at least three zeros in $M$ or $C$, 12 lead to Class Z constraints on $C$, and 24 lead to one diagonal zero in $C$. The remaining 6 structures lead to the Class 9 constraint, $\det M=0$:
\begin{align}
M_D=\begin{bmatrix}
      0 & \times & \times  \\
      0 & \times & 0   \\
      0 & 0 & \times 
   \end{bmatrix}\,,\quad \quad
   \begin{bmatrix}
         0 & \times & \times \\
         0 & 0 & \times   \\
         0 & \times & 0 
      \end{bmatrix}\,,\quad \quad
         \begin{bmatrix}
               0 & \times & 0  \\
               0 & \times & \times   \\
               0 & 0 & \times 
            \end{bmatrix}\,,
\\
\begin{bmatrix}
      0 & \times & 0 \\
      0 & 0 & \times  \\
      0 & \times & \times 
   \end{bmatrix}\,,\quad \quad
   \begin{bmatrix}
         0 & 0 & \times  \\
         0 & \times & \times  \\
         0 & \times & 0 
      \end{bmatrix}\,,\quad \quad
         \begin{bmatrix}
               0 & 0 & \times  \\
               0 & \times & 0   \\
               0 & \times & \times 
            \end{bmatrix}\,.
\end{align}
\end{enumerate}

\subsection{$\bm{N_R=1}$}
For $N_D=6$, of the 84 different structures of $M_D$ in this case, 78 lead to more than three zeros in $M$ or $C$, and 6 lead to one diagonal or off-diagonal zero in $C$.

\newpage

\end{document}